\documentclass[iop, twocolumn, appendixfloats]{emulateapj}
\usepackage{apjfonts}
\usepackage{natbib}
\usepackage{amsfonts}
\usepackage{amsmath}
\usepackage[usenames]{color}
\usepackage{ulem}
\usepackage{verbatim}
\usepackage{longtable}
\usepackage{array}
\usepackage{epsfig}
\usepackage{scrextend}
\usepackage{mathrsfs}
\usepackage{rotating}
\newcommand\msun{{M_\odot}}

\newcommand\teff{T_{\rm eff}}

\newcommand\ionn[2]{#1$\;${\scshape{#2}}}
     
\shorttitle{}
\shortauthors{}

\begin{document}

\title{The Evolution and Properties of Rotating Massive Star Populations}

\author{Jieun Choi\altaffilmark{1}}
\author{Charlie Conroy\altaffilmark{1}}
\author{Nell Byler\altaffilmark{2}}

\altaffiltext{1}{Harvard-Smithsonian Center for Astrophysics, Cambridge, MA 02138, USA}
\altaffiltext{2}{University of Washington, Seattle, WA 98195, USA}

\begin{abstract}
We investigate the integrated properties of massive ($>10~\msun$), rotating, single-star stellar populations for a variety of initial rotation rates ($v/v_{\rm crit}=0.0$, 0.2, 0.4, 0.5, and 0.6). We couple the new MESA Isochrone and Stellar Tracks (MIST) models to the Flexible Stellar Population Synthesis (FSPS) package, extending the stellar population synthesis models to include the contributions from very massive stars ($>100~\msun$), which can be significant in the first $\sim4$~Myr after a starburst. These models predict ionizing luminosities that are consistent with recent observations of young nuclear star clusters. We also construct composite stellar populations assuming a distribution of initial rotation rates. Even in low-metallicity environments where rotation has a significant effect on the evolution of massive stars, we find that stellar population models require a significant contribution from fast-rotating ($v/v_{\rm crit}>0.4$) stars in order to sustain the production of ionizing photons beyond a few Myr following a starburst. These results have potentially important implications for cosmic reionization by massive stars and the interpretation of nebular emission lines in high-redshift star-forming galaxies.
\end{abstract}

\maketitle

\section{Introduction}
Massive stars, though rare in number, are energetically dominant across a wide range of environments spanning star cluster to extragalactic scales. Stellar feedback from young, massive stars in \ionn{H}{ii} regions and star-forming galaxies is a crucial yet poorly understood process. Broadly speaking, stellar feedback refers to the deposition of energy, momentum, mass, and nuclear-burning products via channels that include type I and type II supernovae (SNe), stellar radiation, and winds \citep[e.g.,][]{Murray2010, Hopkins2011}. These processes influence the state of the interstellar medium \citep[e.g.,][]{McKee1977}, thereby regulating star formation \citep[e.g.,][]{Williams1997, MacLow2004, McKee2007} and driving both turbulence and galactic-scale outflows \citep[e.g.,][]{Dekel1986, Martin1999, deAvillez2004, Joung2006, Oppenheimer2006, Agertz2009, Tamburro2009, Hopkins2012, Creasey2013}. 

Observations of star-forming giant molecular clouds (GMCs) have suggested that early feedback processes disperse the dense gas well before the first SNe explode, which may increase the overall efficiency of feedback and reduce the star formation efficiency \citep[e.g.,][]{Evans2009, Krumholz2012a}. Proposed mechanisms of this early feedback include the destruction of the dense molecular regions by expanding \ionn{H}{ii} bubbles \citep[e.g.,][]{Whitworth1979, Matzner2002, Walch2012, Lopez2014} and radiation pressure \citep[e.g.,][]{Murray2005, Krumholz2009}. The complex interplay between the properties of young stellar populations and the dissolution of their birth gas has significant implications for the number of photons that are able to leak out of the host galaxies \citep[e.g.,][]{Dove1994, Gnedin2000, Ma2015} and drive cosmic reionization \citep[e.g., ][]{Haardt1996}.

The influence of stellar feedback in a galactic context has been explored by many groups, both analytically and in simulations \citep[e.g.,][]{Haehnelt1995, Murray2005, Nath2009, Murray2011, Hopkins2011, Hopkins2012, Agertz2013, Ma2015, Muratov2015, Fierlinger2016}. Prescriptions for the time-dependent momentum and energy deposition of stellar populations require stellar population synthesis (SPS) models such as Starburst99 (SB99; \citealt{Leitherer1999, Vazquez2005, Leitherer2010, Leitherer2014}), which in turn rely on stellar evolution models, e.g., the Geneva models \citep{Ekstrom2012, Georgy2013} and realistic stellar spectra, especially in the EUV. Stellar population models that include the effects of stellar rotation lead to an overall increase in feedback efficiency \citep{Levesque2012, Leitherer2014} because rotation makes stars hotter, brighter, and longer-lived \citep[e.g.,][]{Maeder2000}. In other words, rotating stellar populations have higher bolometric luminosities and sustain harder radiation fields (i.e., more ionizing photons) over a longer period of time compared to non-rotating stellar populations.

Binary interaction is another important and complex aspect of massive star evolution. A large fraction of O- and B-type stars are found in binary or higher multiplicity systems, and an estimated $\gtrsim70\%$ of O-type stars are believed to undergo mass exchange, a third of which likely end up in a binary merger product (e.g., \citealt{Chini2012, Sana2012, Sana2013, deMink2014}; but see also \citealt{Kobulnicky2014}, where the authors found that the multiplicity fraction is closer to $\sim55\%$ for orbital periods less than 5000~days, and that it likely depends on the orbital period and separation). Broadly speaking, binarity and rotation have similar effects on stellar evolution. This is a fortunate aspect since large grids of stellar evolution models generally account for only one of these effects due to the sheer size of the parameter space that would otherwise need to be explored (see \citealt{Song2016} for a recent example that accounts for tidal interactions in a binary system of rotating stars). We focus on single star models with rotation in this work and we compare with predictions from the Binary Population and Spectral Synthesis model (BPASS; \citealt{Eldridge2009}).

In this work, we investigate two features of single massive stars in the context of SPS modeling. First, we explore the contribution from very massive stars (VMSs) to the integrated stellar population properties by extending the initial mass function (IMF) upper mass limit from the fiducial value of $100~\msun$ to $300~\msun$. Recent observations of star clusters have suggested the need to include VMSs in models \citep[e.g.,][]{Crowther2016, Smith2016}, though a top-heavy IMF has also been proposed \citep[e.g.,][]{Turner2015}.

Second, we explore the effects of rotation over a range of rotation rates. We build on the previous work by \cite{Levesque2012}, who investigated the effect of stellar rotation on the resulting stellar population properties. Their conclusions were based on a single rotation value (surface velocity set to $40\%$ of the critical, or break-up, velocity, $v/v_{\rm crit}=0.4$), though the authors also considered a composite population by weighting $30\%$ non-rotating and $70\%$ rotating stellar populations. We also investigate the effects of stellar populations harboring a range of rotations rates, which is supported by observations \citep{Huang2010}, and explore their significance in the context of cosmic reionization. One of the distributions considered in this work includes a tail of fast-rotators with $v/v_{\rm crit}>0.4$, which may be of binary origin \citep{RamirezAgudelo2013}.

Both VMSs and a distribution of rotation rates have interesting implications in light of recent observational and theoretical work. \cite{Smith2016} advocated for VMSs in lieu of rapid rotators to resolve the discrepancy between photometric and spectroscopic ages inferred for a young stellar population in NGC~5253. The authors reasoned that large numbers of rapid rotators, though able to explain the observed nitrogen enhancement and ionizing flux, are in tension with the moderate rotational velocities reported in a recent survey of single O-type stars in 30 Doradus \citep{RamirezAgudelo2013}. We explore whether a combination of moderate rotation rates and VMSs can provide a solution to this problem. Furthermore, \cite{Steidel2014} analyzed rest-frame optical spectra of $z\sim2\textrm{--}3$ star-forming galaxies and found that their position in the ``Baldwin, Phillips, and Terlevich'' (BPT) diagram---a nebular emission line diagnostic first introduced in \cite{Baldwin1981}---is offset relative to their $z=0$ counterparts. The authors concluded that this shift could be explained by a harder stellar ionizing radiation field ($T_{\rm blackbody}=50000\textrm{--}60000$~K), higher ionization parameter, and nitrogen-enhanced nebular gas, suggesting rapid stellar rotation or binaries to explain these observations. In subsequent work, \cite{Steidel2016} found that the BPASS binary star models can self-consistently explain the observed line ratios. This, together with the finding that binary models can sufficiently boost the escape fraction of ionizing photons, $f_{\rm esc}$, to explain cosmic reionization \citep{Ma2016b}, are based on a unique feature of binary models: the ability to sustain sufficiently long-lived ($\gtrsim5$~Myr) far-UV stellar radiation field. We investigate whether this can be explained by rotating stellar populations, particularly those with contributions from fast-rotators.

The paper is organized as follows. In Sections~\ref{section:mist_models} and \ref{section:spsm}, we describe the MESA Isochrones and Stellar Tracks (MIST) and Flexible Stellar Population Synthesis (FSPS) models. Next, in Section~\ref{section:results}, we present the results from stellar population modeling, including an investigation of the effects of VMSs, metallicity, and rotation distributions. In Section~\ref{section:discussion}, we discuss the implications of our models in the context of the cosmic reionization and high-redshift star-forming galaxies. We conclude with a summary in Section~\ref{section:conclusions}.

\section{MIST Models}
\label{section:mist_models}
In this section we provide a brief overview of the MIST models, focussing on the details that are most salient to the evolution of massive, rotating stars. For a comprehensive description and summary, the reader is referred to Section 3 and Table 1 in \cite{Choi2016}.

\subsection{Modules for Experiments in Stellar Astrophysics (MESA)}
The stellar evolutionary tracks are computed using MESA,\footnote{http://mesa.sourceforge.net/} an open-source stellar evolution package \citep{Paxton2011, Paxton2013, Paxton2015}. MESAstar, its 1D stellar evolution module, solves the fully coupled Lagrangian equations of stellar structure and composition. Some of the key advantages of MESA include its robust numerical methods, its modular structure that easily enables a user to adapt it to a wide range of problems in stellar astrophysics, and parallelization via OpenMP. The reader is encouraged to consult the original MESA instrument papers and Section 2 in \cite{Choi2016} for more details.

\subsection{Abundances}
All of the models computed for this work are initialized with the solar-scaled abundances and $Z=Z_{\odot, \rm protosolar}=0.0142$ from \cite{Asplund2009}. The helium mass fraction, $Y$, is computed assuming a linear enrichment law from the primordial helium abundance $Y_{\rm p}=0.249$ \citep{Planck2015} to the protosolar helium abundance $Y_{\odot, \rm protosolar}=0.2703$ \citep{Asplund2009}, where $\Delta Y/\Delta Z=1.5$. 

Although $\rm [Fe/H]$ ranges from $-4.0$ to $0.5$ in the full set of grids published on the MIST website,\footnote{http://waps.cfa.harvard.edu/MIST/} we focus on the range $-2.0\leq \rm [Fe/H]\leq0.0$ in this work.

\begin{figure*}
\centering
\includegraphics[width=0.8\textwidth]{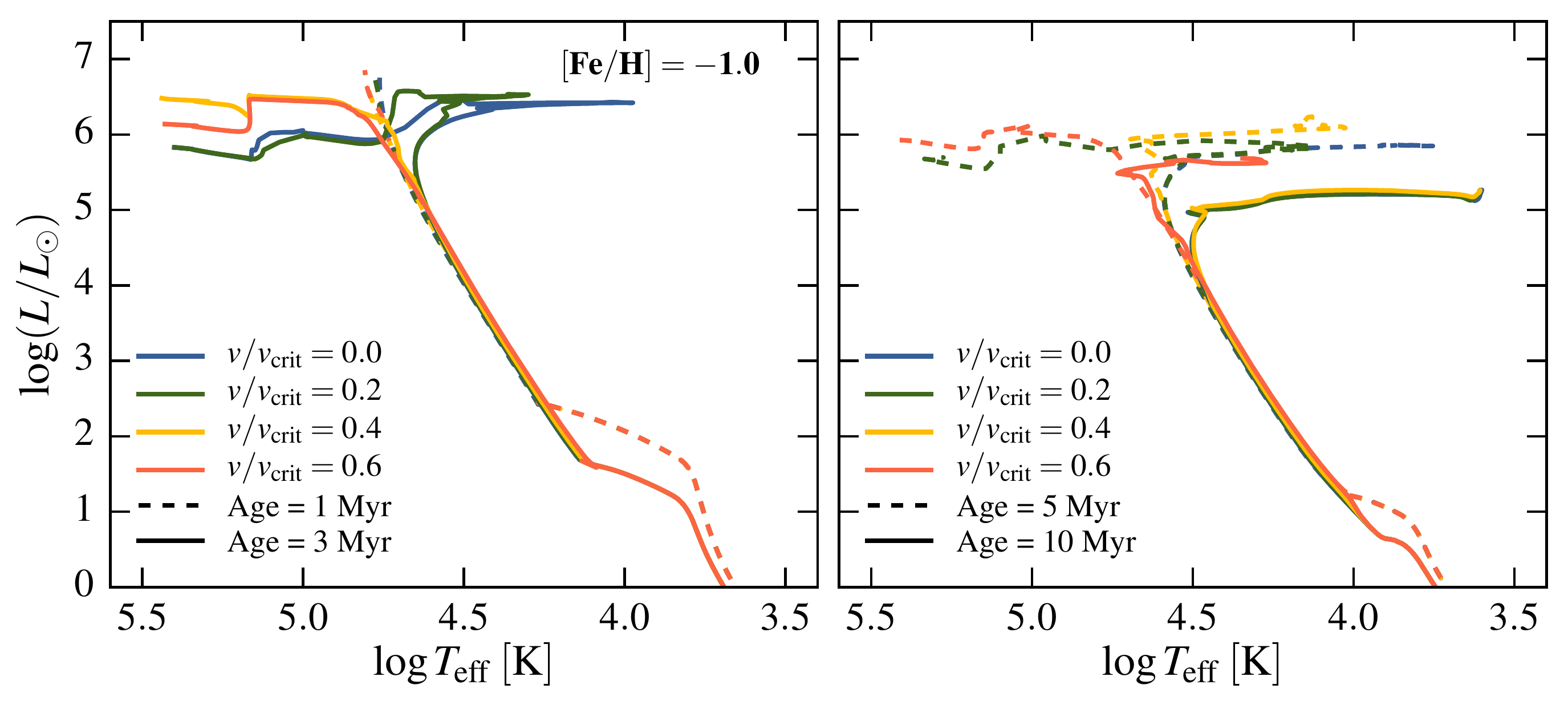}
\caption{$\rm [Fe/H]=-1.0$ isochrones as a function of age and rotation. The four colors correspond to four different values of the initial rotation rates, represented in units of the critical rotation rate. Fast rotation generally leads to hotter, brighter, and more long-lived stars. Left: 1 and 3~Myr. Right:  5 and 10 Myr.}
\label{fig:isochrones_fehm100}
\end{figure*}

\begin{figure*}
\centering
\includegraphics[width=0.8\textwidth]{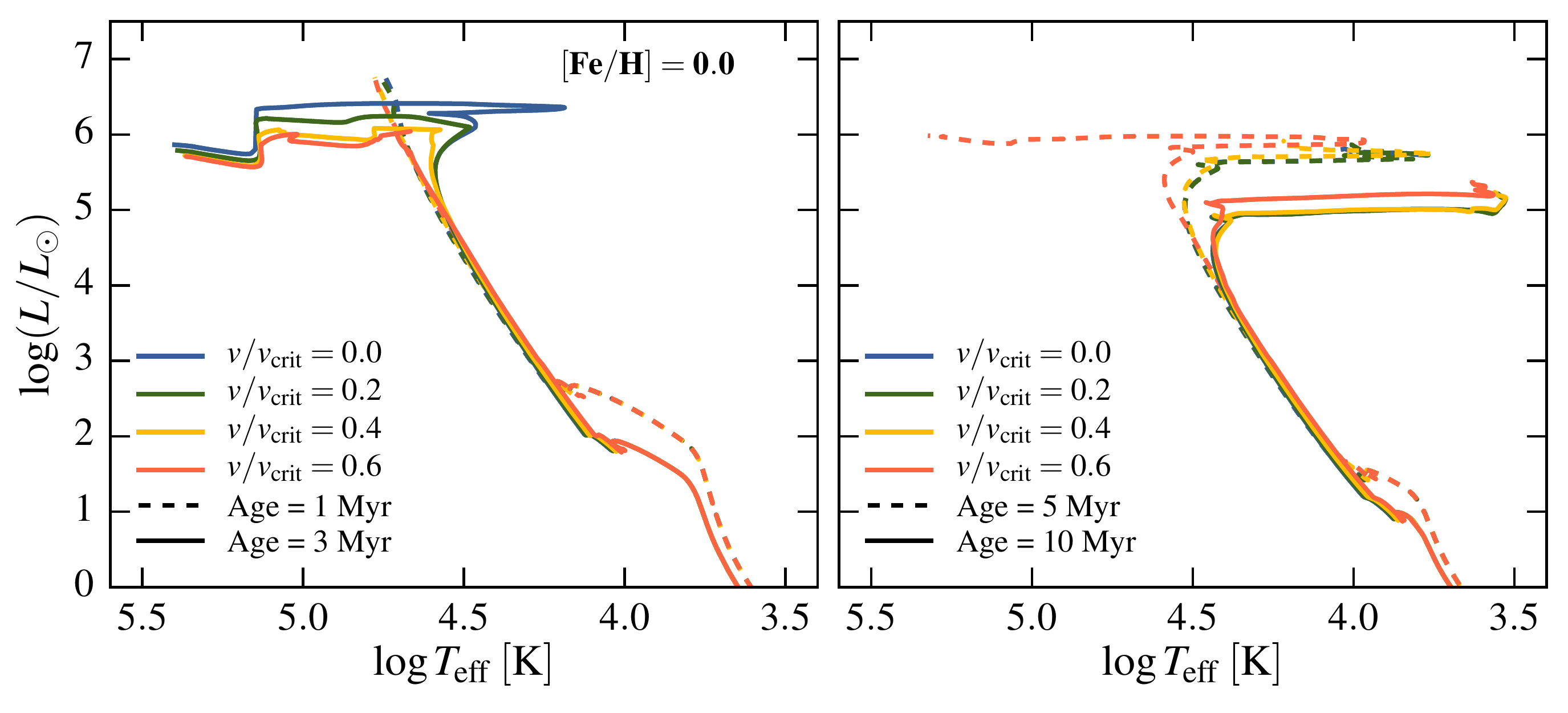}
\caption{Same as Figure~\ref{fig:isochrones_fehm100} except now showing $\rm [Fe/H]=0.0$.}
\label{fig:isochrones_fehp00}
\end{figure*}

\subsection{Rotation}
\label{section:rotation_physics}
The effects of rotation appear in MESA stellar evolution calculations in three main ways. First, rotation decreases the gravitational acceleration $g$ via the centrifugal force, which in turn affects the stellar structure, making the star appear more extended and cooler near the equator. Second, rotation can promote extra mixing in the interior, providing a boost to the transport of chemicals and angular momentum. As a result, the helium fraction---and hence the mean molecular weight $\mu$---is increased in the surface layers and more fuel can be introduced to the convective core, resulting in a hotter, brighter, and more long-lived star. MESA adopts the common approach of treating the chemical and angular momentum transport in a diffusion approximation (\citealt{Endal1978, Zahn1983, Pinsonneault1989, Heger2000} but see also \citealt{Maeder2000, Eggenberger2008, Potter2012a}). Third, rotation enhances mass loss \citep[e.g.,][]{Heger2000, Brott2011, Potter2012a}. MESA adopts the formulation from \cite{Langer1998}, where the mass loss rate $\dot{M}$ is multiplied by a factor that increases dramatically as the surface angular velocity $\Omega$ approaches critical, or break-up, angular velocity, $\Omega_{\rm crit}  = [(1-(L/L_{\rm Edd}))(GM/R^3)]^{0.5}$.\footnote{Note that this expression assumes that the surface brightness is uniform over the stellar surface. \cite{Maeder2000} worked out an alternative expression that takes into account gravity darkening effects, but the difference is small unless $L/L_{\rm Edd}$ is much larger than $\sim0.65$.} Because $\Omega_{\rm crit}$ depends on the ratio of the bolometric luminosity to the Eddington luminosity, $L/L_{\rm Edd}$, massive stars near the Eddington limit can experience a sizable increase in their mass loss rates even with small amount of rotation, resulting in a prompt removal of their surface hydrogen layer \citep[e.g.,][]{Maeder2000, Choi2016}.

Models are initialized following solid-body rotation near the zero-age main sequence (ZAMS), similar to the approach adopted by many other stellar evolution codes \citep{Pinsonneault1989, Heger2000, Eggenberger2008}. We compute models with $v/v_{\rm crit}=0.0$, 0.2, 0.4, 0.5, and 0.6 at ZAMS, where $v/v_{\rm crit}$ corresponds to the ratio of the stellar rotation velocity to the critical rotation velocity at the surface and $v_{\rm crit}  = [(1-(L/L_{\rm Edd}))(GM/R)]^{0.5}$.

\begin{figure*}
\centering
\includegraphics[width=0.8\textwidth]{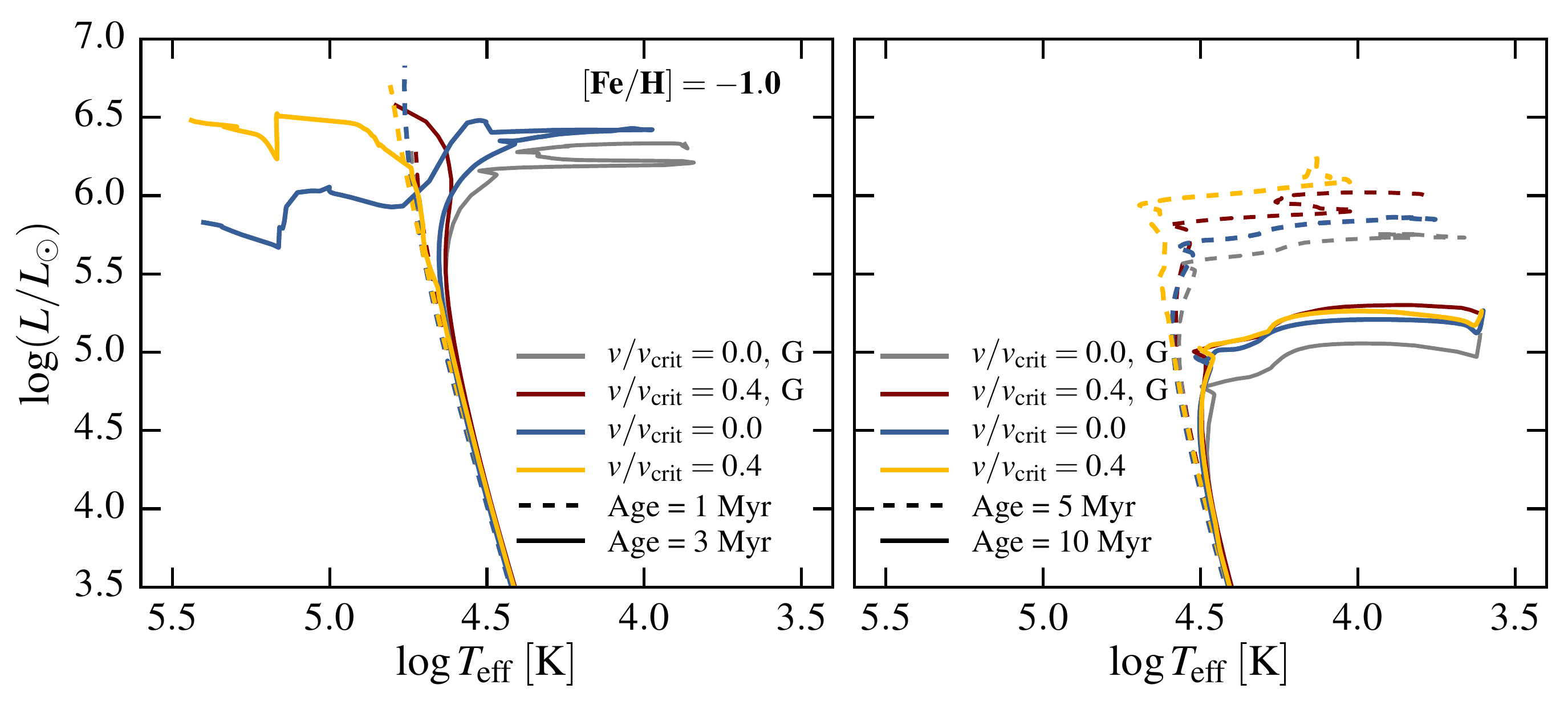}
\caption{Same as Figure~\ref{fig:isochrones_fehm100} except now zooming in near the MSTO. Isochrones for the Geneva rotating and non-rotating models are also included for comparison. For MIST, we only show the $v/v_{\rm crit}=0.0$ and 0.4 isochrones for display purposes.}
\label{fig:isochrones_fehm100_zoom}
\end{figure*}

\begin{figure*}
\centering
\includegraphics[width=0.8\textwidth]{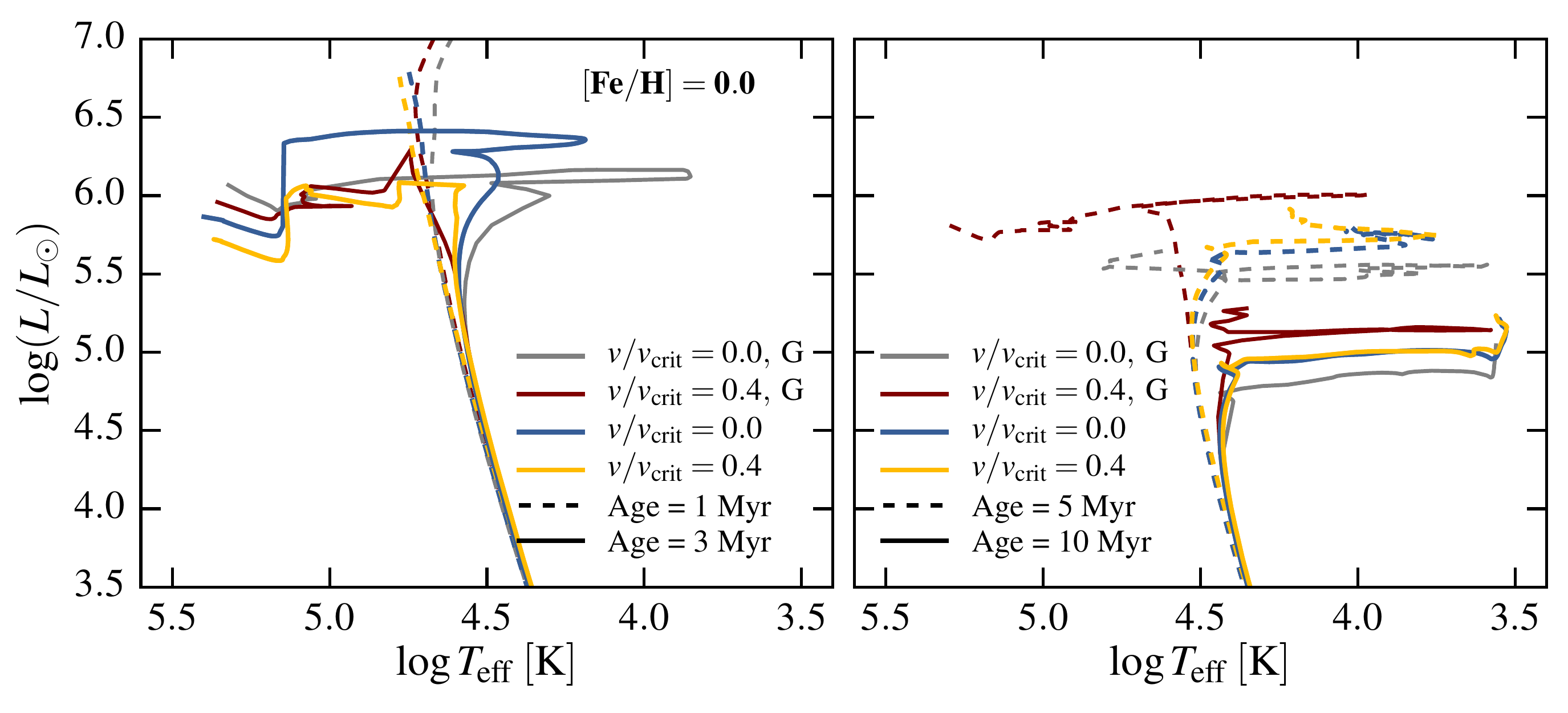}
\caption{Same as Figure~\ref{fig:isochrones_fehm100_zoom} except now showing $\rm [Fe/H]=0.0$.}
\label{fig:isochrones_fehp000_zoom}
\end{figure*}

\subsection{Mixing}
Convective mixing of elements is treated as a time-dependent diffusive process with the diffusion coefficient provided by the mixing length formalism of \cite{Henyey1965} and assuming $\alpha_{\rm MLT}=1.82$ obtained from the solar calibration \citep{Choi2016}. Convectively unstable zones are identified according to the Ledoux criterion, which is similar to the Schwarzschild criterion but also includes composition effects. The effects of convective overshoot mixing beyond the fiducial Ledoux boundary are taken into account following \cite{Freytag1996} and \cite{Herwig2000}, which assumes that the diffusion coefficient decays exponentially with distance from the convective boundary. A free parameter that controls the efficiency of this extra mixing in the core, $f_{\rm ov}=0.016$, is empirically calibrated to reproduce the main-sequence turnoff (MSTO) morphology in M67 (\citealt{Choi2016}, but see also \citealt{Magic2010}). Semiconvection and thermohaline mixing are also included, where we adopt the efficiency parameters $\alpha_{\rm sc}=0.1$ and $\alpha_{\rm th}=666$. Finally, we account for five rotationally induced instabilities that lead to chemical and angular momentum transport, namely the dynamical shear instability, secular shear instability, Solberg-H{\o}iland instability, Eddington-Sweet circulation, and Goldreich-Schubert-Fricke instability \citep{Heger2000, Paxton2013}. We do not include mixing due to internal magnetic fields generated by a Tayler-Spruit dynamo \citep{Spruit2002}.

\subsection{Mass Loss}
In hot stars, mass loss is driven by the momentum transfer from ultraviolet photons to metal ions in the atmosphere \citep{Lucy1970, Castor1975} and is therefore metallicity-dependent. Mass loss in massive stars is separated into three categories in the MIST model calculations. For stars with $\teff > 1.1\times10^4$~K and surface hydrogen mass fraction $X_{\rm surf} > 0.4$, we adopt the mass loss prescription from \cite{Vink2000, Vink2001}. If the star loses a considerable amount of its outer hydrogen layer ($X_{\rm surf} < 0.4$) and becomes a Wolf-Rayet (WR) star, we use the wind prescription of \cite{Nugis2000} instead. Lastly, we use the \cite{deJager1988} prescription for all stars that have effective temperatures below $10^4$~K, including red supergiant (RSG) stars. As discussed in \ref{section:rotation_physics}, mass loss is enhanced in rotating models by a factor that depends on the rotation rate and the Eddington ratio.

\begin{figure*}
\centering
\includegraphics[width=0.8\textwidth]{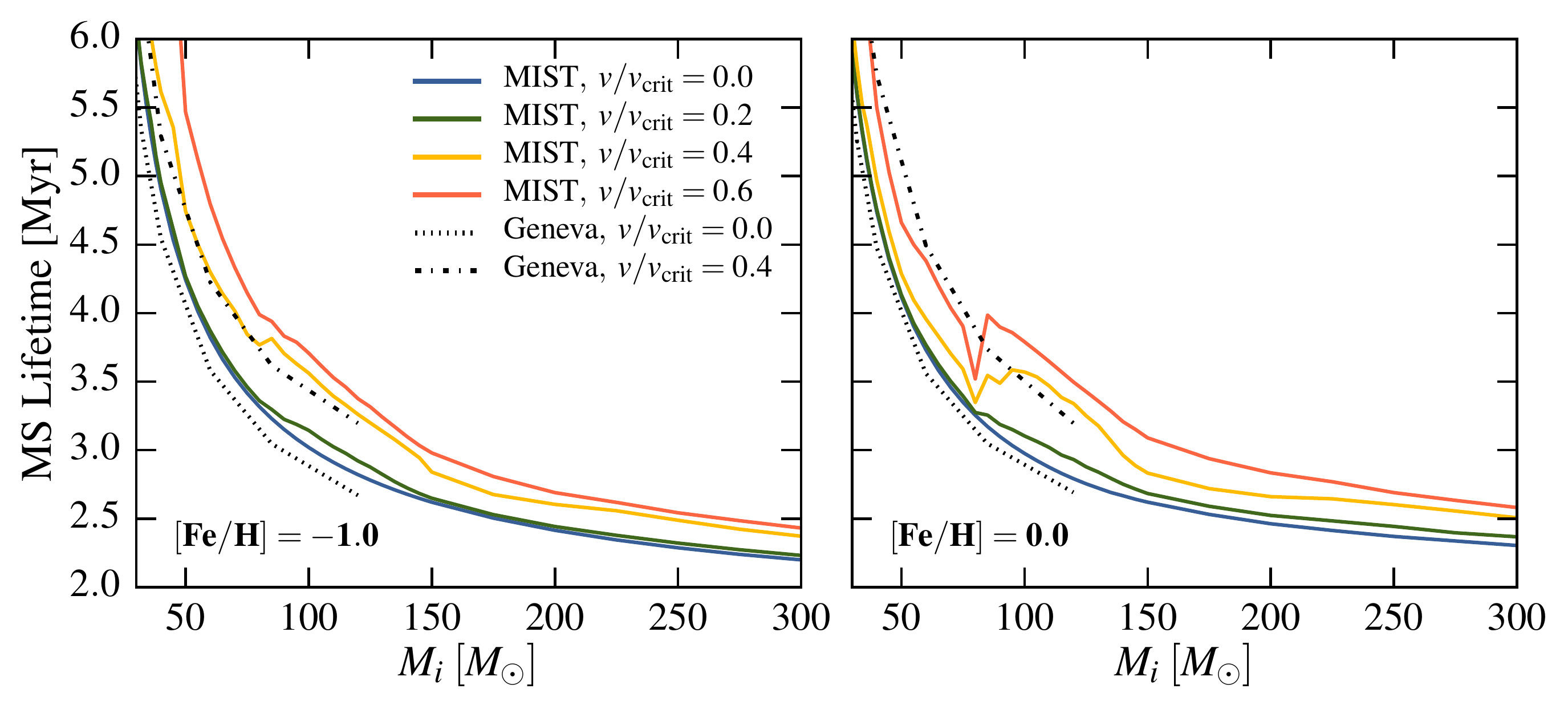}
\caption{Left: MS lifetimes as a function of initial mass for four different values of initial rotation rates at $\rm [Fe/H]=-1.0$. The black curves represent MS lifetimes for the Geneva \citep{Ekstrom2012, Georgy2013} evolutionary tracks at $Z=0.002$ ($\rm [Fe/H] = -0.86$ assuming their adopted $Z_{\odot} = 0.014$). Right: same as the left panel except now showing $\rm [Fe/H]=0.0$.}
\label{fig:ms_lifetimes}
\end{figure*}

\subsection{Tracks and Isochrones}
We compute MIST isochrones for five rotation rates---$v/v_{\rm crit}=0.0, 0.2, 0.4, 0.5$, and $0.6$---for each metallicity. The stellar evolutionary tracks ranging from 0.1 to $300~\msun$ are computed by MESA and processed into isochrones following the procedure outlined in \cite{Dotter2016}. We modified the mass sampling in the isochrones in this work compared to that in the standard MIST grids to ensure that fast evolutionary phases at early times are particularly well-represented. Figures~\ref{fig:isochrones_fehm100} and \ref{fig:isochrones_fehp00} show 1, 3, 5, and 10~Myr isochrones at $\rm [Fe/H]=-1.0$ and $0.0$, respectively. We omit $v/v_{\rm crit}=0.5$ from these figures for display purposes. There are several notable features. First, the rotating models tend to be hotter and more luminous overall. Second, WR stars from very massive progenitors appear between 1~Myr and 5~Myr, particularly for $v/v_{\rm crit}>0.2$. Third, some isochrones, e.g., $v/v_{\rm crit}=0.6$ at $\rm [Fe/H]=-1.0$, show signatures of quasi-chemically homogeneous evolution \citep[QCHE;][]{Maeder1987, Langer1992}. QCHE occurs in massive, fast-rotating stars where the mixing timescale becomes shorter than or comparable to the nuclear burning timescale. In this scenario, nuclear burning products---mostly helium---from the core are mixed into the outer layers and a large fraction of the star undergoes nuclear fusion as fresh fuel is channeled into the core. Overall, the star becomes hotter due to the reduced mean opacity, and brighter and more long-lived due to enhanced mixing in the core (see Figures 5, 6, and 7 in \citealt{Brott2011} for examples). QCHE occurs more readily at low metallicity due to the reduced angular momentum loss via stellar winds \citep{Yoon2005, Woosley2006} and the more compact stellar structure resulting in a decreased mixing timescale ($\tau \sim R^2/D$, where $R$ is the radius of the star and $D$ is the diffusion coefficient for mixing). To verify that both processes indeed contribute, we performed a simple test where we ran a total of eight $60~\msun$ models, with and without rotation, with and without mass loss, at solar and one-tenth-solar metallicities. We found that even in the absence of mass loss, i.e., no angular momentum loss, the MS lifetime enhancement due to rotation was still larger for the low metallicity model ($\sim20\%$) compared to the solar metallicity model ($\sim10\%$), which suggests that the compactness of the star at low metallicity is indeed an important effect. In the rotating case, there was a small ($\lesssim5\%$) enhancement in the MS lifetime for the models without any mass loss compared to those with mass loss. An extensive study examining the relative importance of the two effects as a function of metallicity, rotation rate, mass, etc. is beyond the scope of this paper, but would be worth exploring further.

In Figures~\ref{fig:isochrones_fehm100_zoom} and \ref{fig:isochrones_fehp000_zoom} we compare the $v/v_{\rm crit}=0.0$ and 0.4 MIST isochrones with the non-rotating and rotating Geneva isochrones, zooming in around the MSTO for clarity. In detail, there are clear morphological differences between the MIST and Geneva isochrones, but both their evolution with time and their modifications due to rotation are qualitatively similar. A notable difference is the magnitude of the effect of rotation on the MS lifetimes and the trajectory in the HR diagram at $t\gtrsim5$~Myr; rotating Geneva models are much brighter than the non-rotating models compared to their MIST counterparts.

In order to investigate the effect of rotation on the MS lifetimes of massive stars in more detail, we plot the MS lifetime--initial mass relations in Figure~\ref{fig:ms_lifetimes}. The black dotted and dot-dashed lines correspond to the non-rotating and rotating Geneva \citep{Ekstrom2012, Georgy2013} evolutionary tracks for $Z=0.002$ ($\rm [Fe/H] = -0.86$ assuming their adopted $Z_{\odot} = 0.014$) and $Z=0.014$. At a fixed initial mass, a higher rotation rate lengthens the MS lifetime due to more efficient rotational mixing. Note that these plots demonstrate that the MS lifetime boost at a fixed initial mass is larger in the Geneva models compared to that in the MIST models (see also Figure 20 in \citealt{Choi2016}), suggesting that rotational mixing may be more efficient in the former. The MS lifetime flattens out at the high mass end since $t \propto L/M$ and luminosity becomes a shallower function of mass: for $L\propto M^\alpha$, we find $\alpha\sim4$ for $M\lesssim10~\msun$ and $\alpha\sim1.5$ for $M\gtrsim100~\msun$. At solar metallicity, there is an interesting non-monotonic behavior around $80~\msun$. Rapidly rotating stars with masses greater than $80~\msun$ begin to evolve more vertically and blueward in the HR diagram during the MS.\footnote{Although quasi-chemically homogeneous evolution appears more readily in low-metallicity environments, they may also occur in metal-rich systems. See \cite{Martins2013} for more details.} Due to their high luminosities and temperatures and therefore heavy mass loss, they quickly shed their H-rich envelopes and become WR stars. As they continue to lose significant mass due to heavy WR winds, they effectively ``reset'' as lower-mass stars, which prolongs their MS lifetimes. In other words, the MS lifetime--initial mass relation systematically shifts upward to another branch beyond $80~\msun$.

\begin{figure*}
\centering
\includegraphics[width=0.9\textwidth]{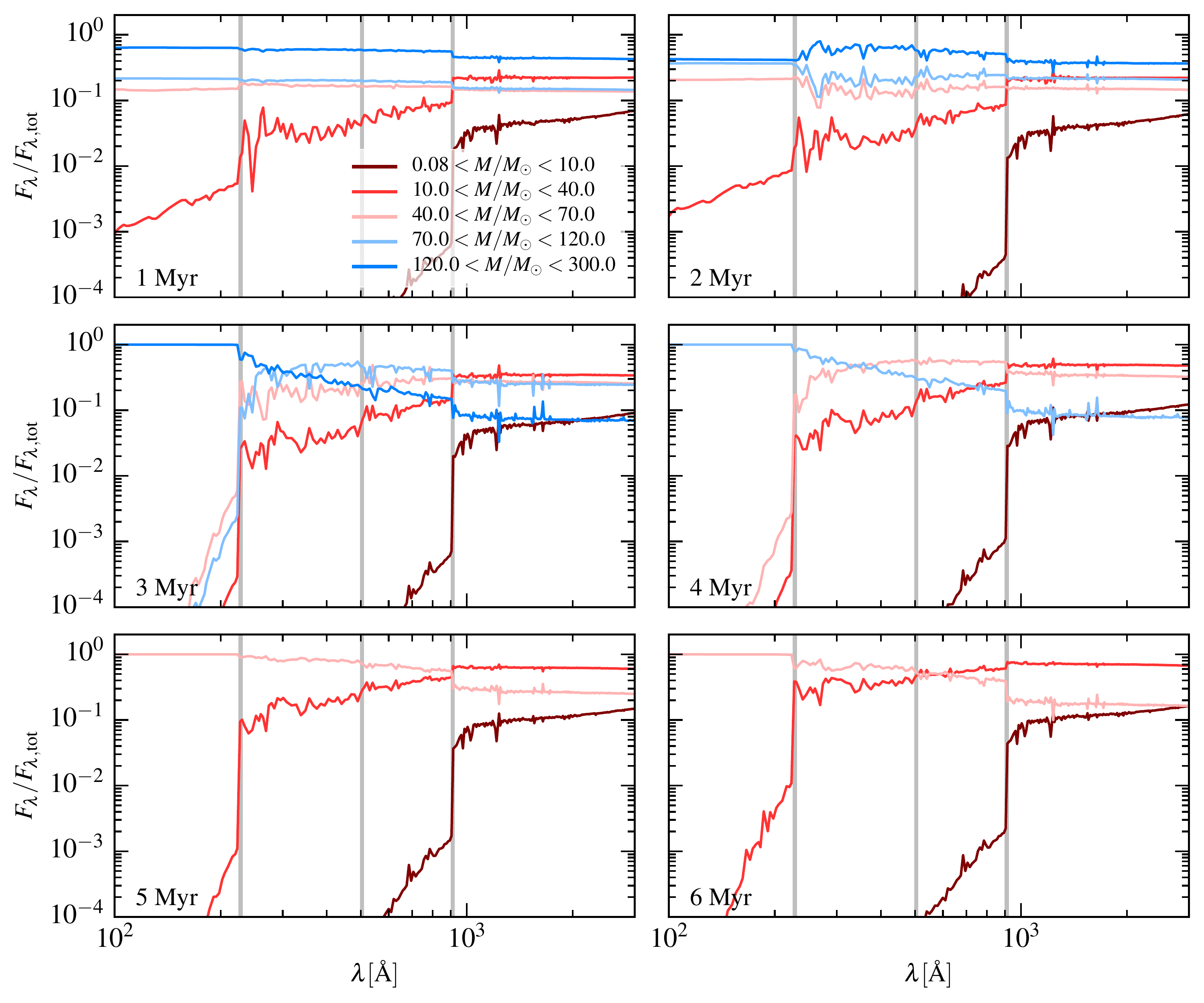} 
\caption{Fractional flux contribution to the total ultraviolet flux from stars in different mass ranges for a single-burst population with $v/v_{\rm crit}=0.6$ at $\rm [Fe/H]=-1.0$ over 1\textrm{--}6~Myr. The three grey vertical lines mark the wavelengths below which photons can ionize hydrogen, singly ionize helium, and doubly ionize helium (912~\AA, 504~\AA, and 228~\AA).} 
\label{fig:feh_zmet1_ionizing_photons_imf300}
\end{figure*}

\section{Stellar Population Synthesis Models}
\label{section:spsm}
\subsection{Flexible Stellar Population Synthesis}
\label{section:fsps}
Models of simple stellar populations (SSPs) in this work are computed using the Flexible Stellar Population Synthesis package \citep[FSPS;][]{Conroy2009, Conroy2010}. The primary stellar spectral library consists of the MILES empirical library \citep{SanchezBlazquez2006, FalconBarroso2011}, which is supplemented with the CMFGEN WR spectra (available for both WN and WC subtypes) from \cite{Smith2002} and the WM-Basic \citep{Pauldrach2012} hot star spectra (J.J. Eldridge, priv. comm.). The WR spectra are assigned to points in the isochrone that are first identified as WR then further categorized into WR subtypes, e.g., WN and WC, according to the surface composition: stars with surface C/O ratio $>1$ and $\leq1$ are labeled as WC and WN, respectively. The WM-Basic models are applied for MS stars with $T_{\rm eff}>2.5\times10^4~\rm K$.

For all subsequent analyses, we simulate a population of total mass $10^6~\msun$ following an instantaneous burst of star formation. All integrated quantities are computed following the \citealt{Kroupa2001} initial mass function (IMF), assuming that the IMF is fully sampled. The IMF lower and upper mass limits are set to $0.08~\msun$ and $300~\msun$, respectively (see Section~\ref{section:imf_upper_mass_limit} for a discussion). Stochastic IMF sampling effects are likely to have a bearing on real stellar populations, especially in low-mass systems \citep[e.g.,][]{Cervino2000, Cervino2001, daSilva2012, daSilva2014}. We revisit this point briefly in Section~\ref{section:imf_stochasticity}.

As discussed in \cite{Maeder1990}, assigning a spectrum to a WR star in the isochrone is a non-trivial task due to the high optical depth of its wind. As a result, the effective photosphere---the $\tau=2/3$ surface---has a radius larger than the hydrostatic radius reported in the stellar evolutionary track, and the inferred temperature from observations, $T_{\rm eff,\;WR}$, is cooler than the temperature from stellar evolution calculations, $T_*$. We follow the standard approach adopted by \cite{Maeder1990} and \cite{Smith2002}, where $T_{\rm eff,\;WR}$ is estimated via a weighted sum of $T_*$ and $T_{\rm eff,\;wind}$. $T_{\rm eff,\;wind}$ is computed by assuming a velocity law:

\begin{equation}
v(r) = v_{\infty} \left ( 1 - \frac{R}{r} \right)^2\;\;,
\end{equation}
where $v_{\infty}$ is the terminal velocity, set to $3\times10^8~\rm cm\;s^{-1}$.

Starting with the definition of optical depth $d\tau = -\kappa \rho \;dr$ and integrating inward to $\tau=2/3$, the effective photosphere is obtained

\begin{equation}
R_{\rm eff,\;WR} = R_{*} + \frac{3\kappa |\dot{M}|}{8\pi v_{\infty}}\;\;.
\end{equation}

Since $L$ is constant, it is trivial to compute $T_{\rm eff,\;WR}$ in terms of $R_{\rm eff,\;WR}$, $R_{*}$, and $T_{*}$. Then the temperature used to retrieve the spectrum for a WR star with $T_*$ is $0.6T_* + 0.4T_{\rm eff,\;WR}$. \cite{Smith2002} chose this weighting scheme to reproduce the range of observed temperatures of the Galactic and LMC WR population. In principle, the weighting factors should be re-derived for this work given the differences in the underlying stellar evolutionary models adopted in \cite{Smith2002} (\citealt{Meynet1994} tracks) and MIST. However, the objective here is a simple assessment of the uncertainties introduced due to the ill-defined WR temperatures. We revisit this point in Section~\ref{section:ionizing_photons}.

\begin{figure*}
\centering
\includegraphics[width=\textwidth]{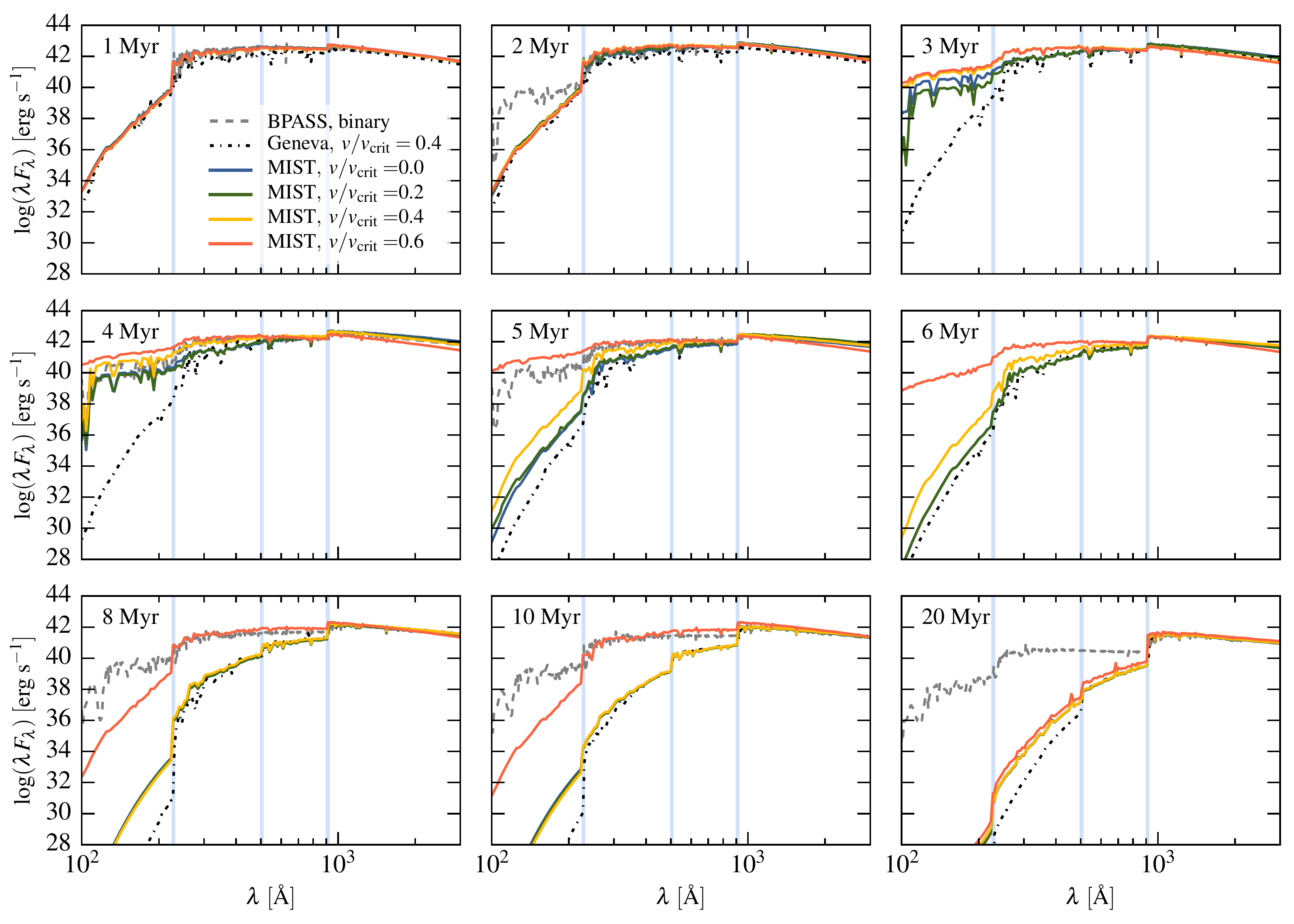}
\caption{Time evolution of SED model predictions for a single-burst, $10^6~\msun$ stellar population at $\rm [Fe/H]=-1.0$. There are four initial rotation rates considered for FSPS+MIST, a binary-star model from BPASS, and a rotating model from SB99+Geneva. The $v/v_{\rm crit}=0.5$ model is excluded for clarity. Note that BPASS is not shown in every panel. The three blue vertical lines mark the wavelengths of photons able to ionize hydrogen, singly ionize helium, and doubly ionize helium (912~\AA, 504~\AA, and 228~\AA). The FSPS+MIST and BPASS models assume an IMF upper mass limit of $300~\msun$ while the SB99+Geneva model assumes $100~\msun$. Only the $v/v_{\rm crit}=0.6$ and binary models are capable of producing significant EUV flux for $\rm \geq 5~Myr$ and only the binary model can produce substantial EUV flux beyond 10 Myr.} 
\label{fig:feh_m1_sed}
\end{figure*}

\begin{figure*}
\centering
\includegraphics[width=\textwidth]{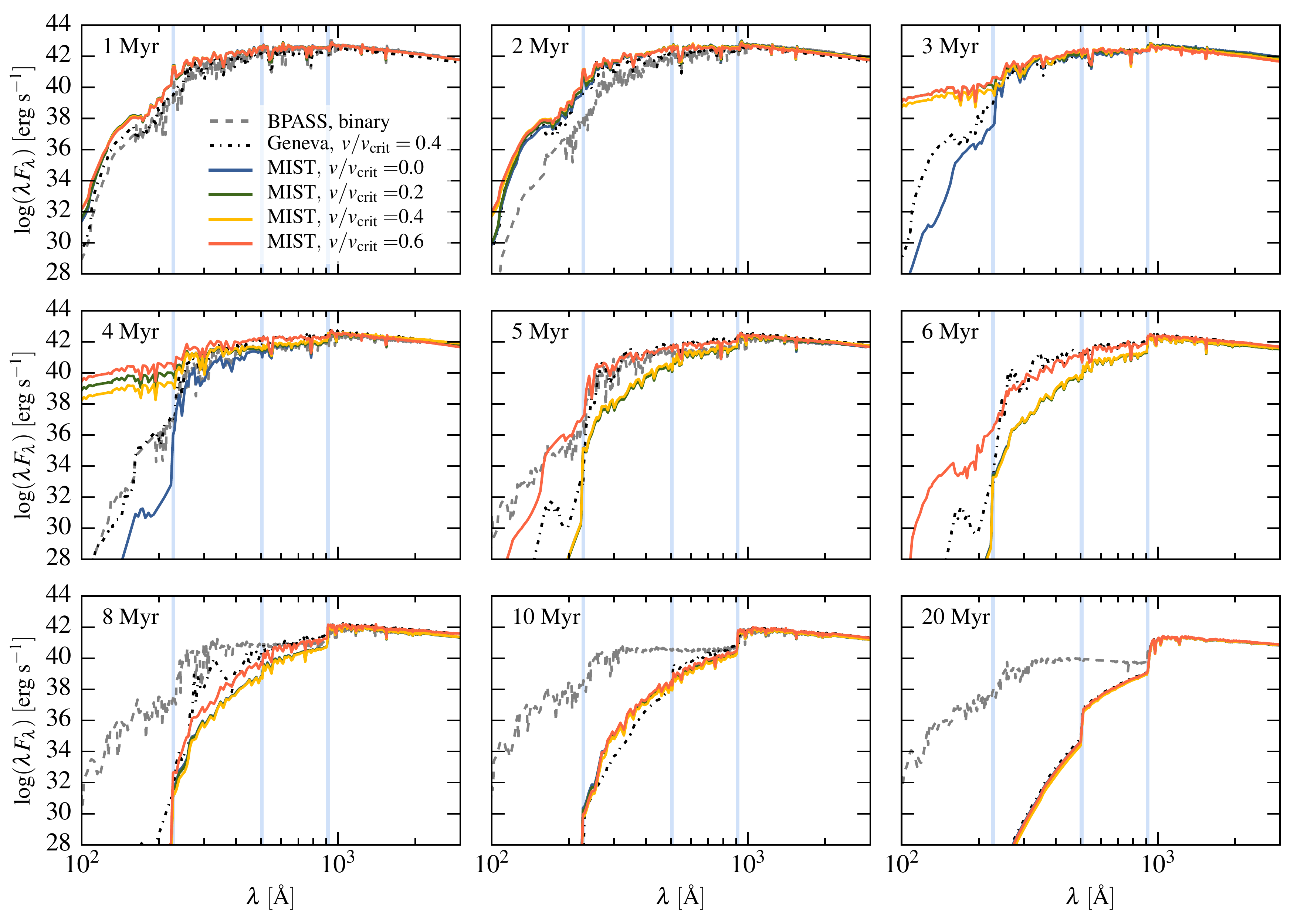}
\caption{Same as Figure~\ref{fig:feh_m1_sed} except now showing $\rm [Fe/H]=0.0$.} 
\label{fig:feh_p0_sed}
\end{figure*}

\subsection{IMF Stochasticity}
\label{section:imf_stochasticity}
We evaluate the assumption that the IMF is smoothly sampled for this work. A key concern regarding IMF stochasticity in the context of SPS modeling is that many quantities of interest, e.g., ionizing flux, are dominated by a very small number of the most massive stars, a subpopulation that is also the rarest and thus most prone to sampling effects. Thus two stellar populations that are otherwise completely identical in parameter space (e.g., metallicity, total mass, age) may appear quite different depending on the stellar mass distribution. Figure~\ref{fig:feh_zmet1_ionizing_photons_imf300}, which shows the fractional flux contribution from stars in different mass ranges, provides a sense of how much variation there may be in spectral energy distributions (SEDs) due to the IMF sampling effects.

There are alternative approaches to SPS modeling \citep[e.g.,][]{Barbaro1977, Cervino2000, Villaverde2010, Eldridge2012, daSilva2012, daSilva2014} that simulate the IMF stochastic sampling effects. \cite{Cervino2002} constructed a statistical formalism to quantify the uncertainties due to IMF sampling errors and applied their results to model a starburst population from birth to $20~\rm Myr$. Overall, they found that a minimum total stellar mass of $10^5~\msun$ is necessary to ensure a relative dispersion of less than 10\% in hydrogen and helium ionizing flux, though during the ``WR phase'' (between 2 and 5 Myr), this threshold mass is increased to $10^6~\msun$. 

Since we are operating under the assumption that the total stellar mass of the cluster modeled in our work is $10^6~\msun$, random IMF sampling effects are likely unimportant for our conclusions. However, the exact value of this critical cluster mass depends on the adopted IMF (e.g., functional form, lower and upper mass limits). The conclusions from \cite{Cervino2002} were based on a \citealt{Salpeter1955} IMF over the mass range $2\textrm{--}120~\msun$. Even though the increased mass range ($0.08\textrm{--}300~\msun$) in our work would push the threshold total mass upward, an ensemble of such clusters in e.g., a starburst galaxy, should collectively smear out the sampling effects \cite[see e.g.,][]{Cervino2000, Eldridge2012}.

\begin{figure}
\centering
\includegraphics[width=0.45\textwidth]{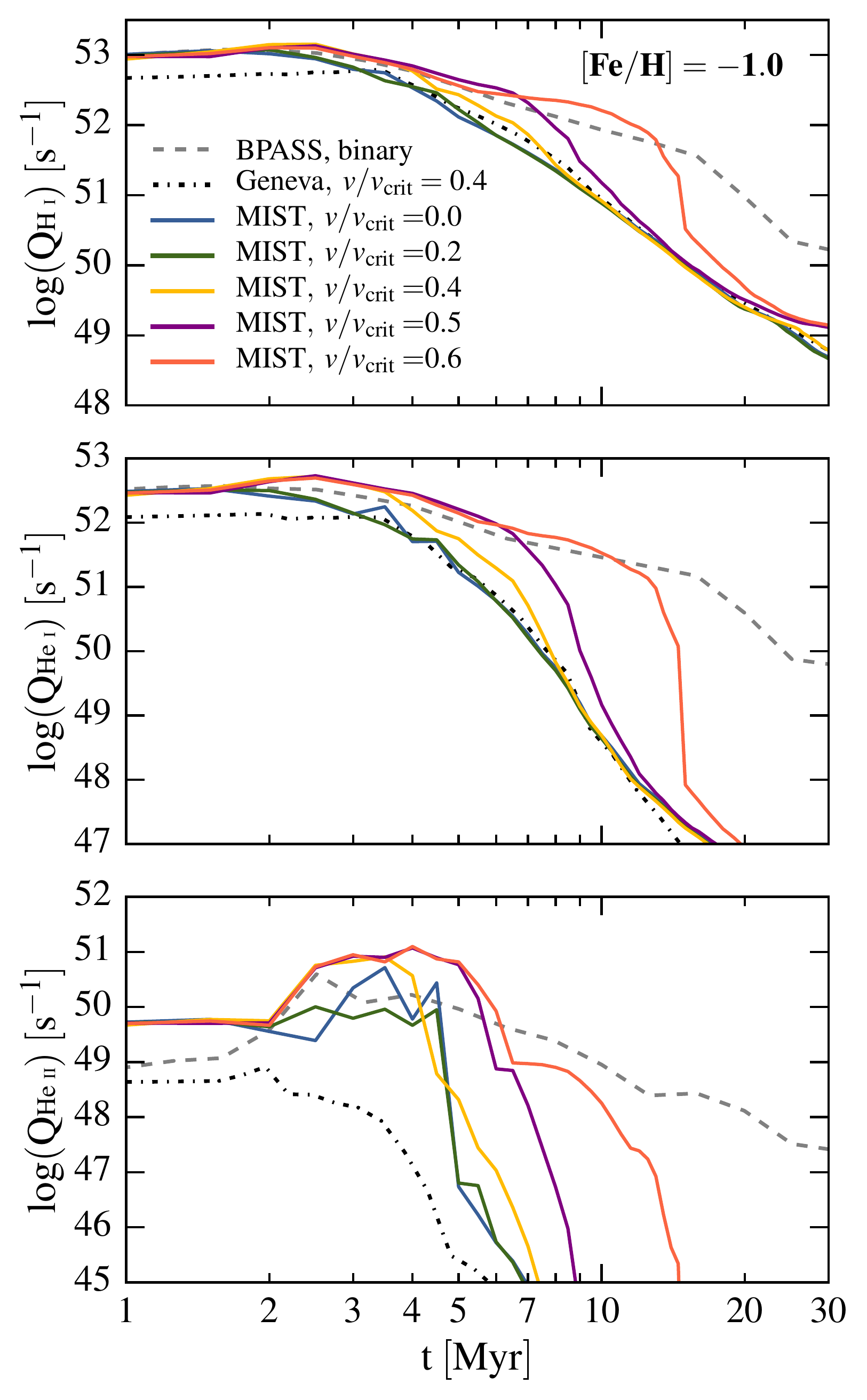} 
\caption{Time evolution of the ionizing photon luminosity from the FSPS+MIST, BPASS, and SB99+Geneva models for a $10^6~\msun$ single-burst stellar population at $\rm [Fe/H]=-1.0$. The FSPS+MIST and BPASS models assume an IMF upper mass limit of $300~\msun$ while the SB99+Geneva model assumes $100~\msun$, which explains the discrepancy at the earliest ages. Non-rotating and moderately rotating models predict a steep decline in the ionizing luminosity with time. Top: photons capable of ionizing hydrogen. Middle: photons capable of singly ionizing helium. Bottom: photons capable of doubly ionizing helium.}
\label{fig:feh_m1_ionizing_photons}
\end{figure}

\begin{figure}
\centering
\includegraphics[width=0.45\textwidth]{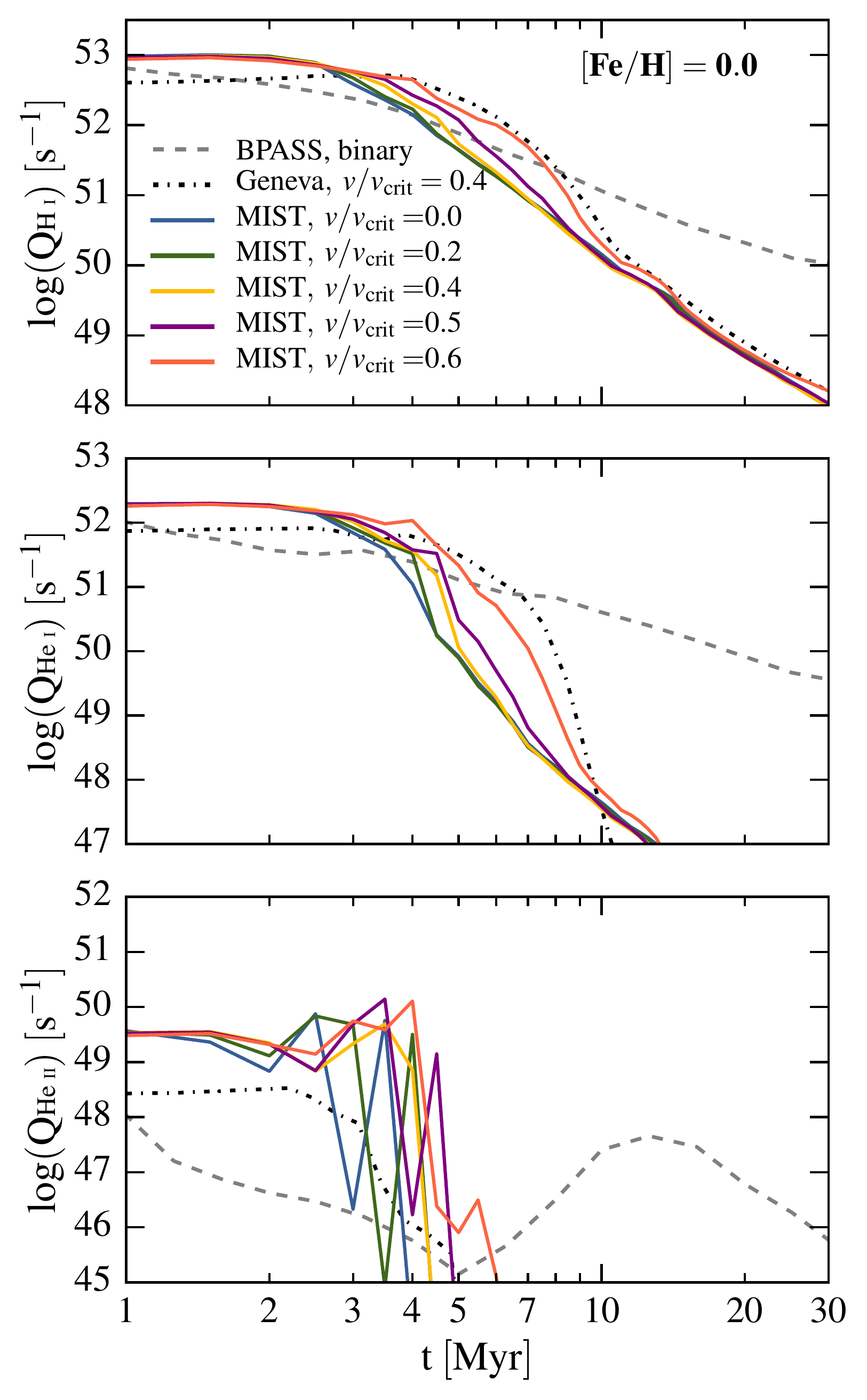} 
\caption{Same as Figure~\ref{fig:feh_m1_ionizing_photons} except now showing $\rm [Fe/H]=0.0$.} 
\label{fig:feh_p0_ionizing_photons}
\end{figure}

\begin{figure}
\centering
\includegraphics[width=0.45\textwidth]{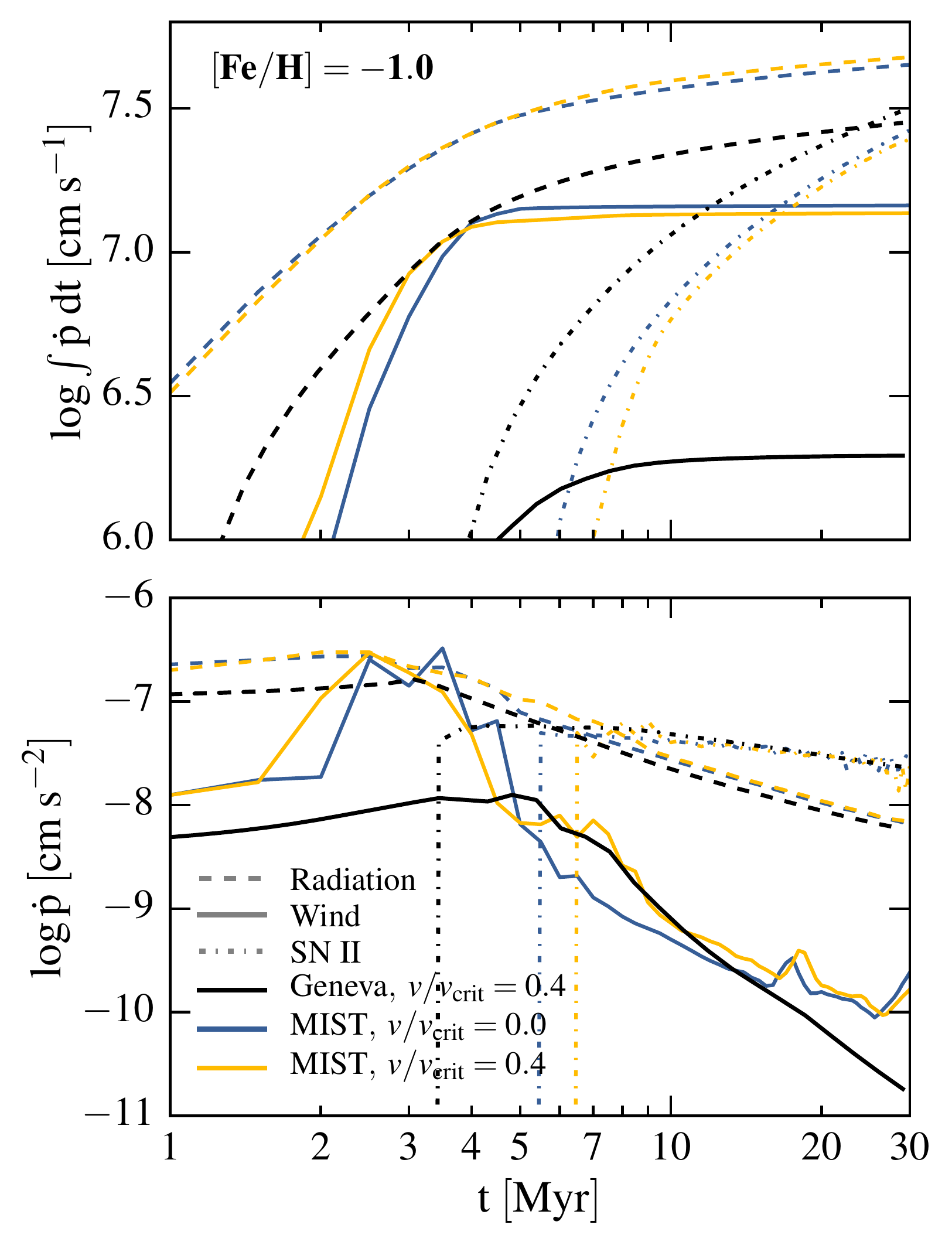}
\caption{Top: cumulative specific momentum injected via radiation, wind, and type II supernovae for a single-burst stellar population at $\rm [Fe/H]=-1.0$. We compare $v/v_{\rm crit}=0.0$ (blue) and 0.4 (yellow) FSPS+MIST model predictions with that from SB99+Geneva, which is shown as a black curve. We assume a $\tau_{\rm IR} = 1$ for simplicity. Bottom: instantaneous specific momentum. The FSPS+MIST models assume an IMF upper mass limit of $300~\msun$ while the SB99+Geneva model assumes $100~\msun$. This differences largely explains the offset between the two models at very early times.} 
\label{fig:feh_m1_feedback}
\end{figure}

\begin{figure}
\centering
\includegraphics[width=0.45\textwidth]{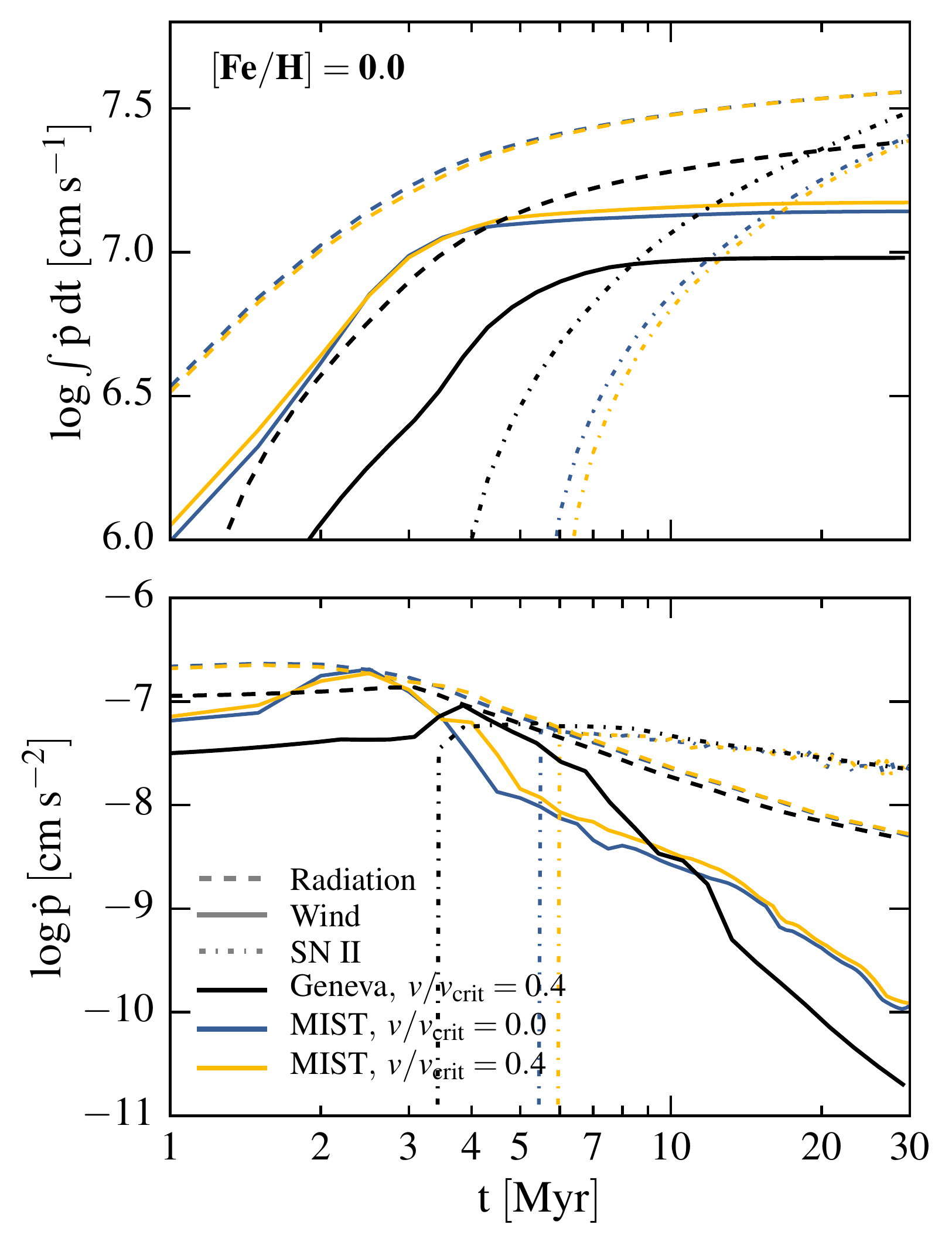}
\caption{Same as Figure~\ref{fig:feh_m1_feedback} except now showing $\rm [Fe/H]=0.0$.} 
\label{fig:feh_p0_feedback}
\end{figure}

\vspace{0.5cm}
\section{Stellar Population Properties}
\label{section:results}
\subsection{Spectral Energy Distributions (SEDs)}
Figures~\ref{fig:feh_m1_sed} and \ref{fig:feh_p0_sed} show SEDs from 1 to 20~Myr for $\rm[Fe/H]=-1.0$ and 0.0, respectively. The colored curves represent four different initial rotation rates for FSPS+MIST, while the gray and black curves show the BPASS v2.0\footnote{IMF slopes of $-1.30$ for $0.1<M_i/\msun<0.5$ and $-2.35$ for $0.5<M_i/\msun<300$.} \citep{Stanway2016} and SB99+Geneva \citep{Leitherer2014} predictions\footnote{IMF slopes of $-1.3$ for $0.1<M_i/\msun<0.5$ and $-2.3$ for $0.5<M_i/\msun<100$.} for comparison where available. The $v/v_{\rm crit}=0.5$ FSPS+MIST model is excluded for clarity. The three blue vertical lines mark the wavelengths of photons capable of ionizing hydrogen, singly ionizing helium, and doubly ionizing helium (912~\AA, 504~\AA, and 228~\AA). For clarity, we do not include the BPASS single star or SB99+Geneva non-rotating models, but they are in broad agreement with the FSPS+MIST non-rotating models.

The FSPS+MIST model predicts a harder spectrum compared to the SB99+Geneva model, partially due to differences in the underlying isochrones. Moreover, the FSPS+MIST model presented here is slightly more metal-poor than the SB99+Geneva model. The large discrepancy at 3~Myr is due to the lack of stars more massive than $300~\msun$ in the SB99+Geneva models.\footnote{The Geneva group has published stellar evolutionary tracks of VMSs (120 to $500~\msun$) at solar, LMC, and SMC metallicities \citep{Yusof2013}, but the corresponding isochrones are not incorporated into SB99 at the time of writing.} With the exception of $v/v_{\rm crit}=0.6$, the FSPS+MIST models cease to produce an appreciable amount of photons blueward of 228~\AA{} beyond $\sim6$~Myr, while the BPASS binary models continue to output significant EUV flux even at very late times (20~Myr). This can be understood by recalling that single star models rely exclusively on the most massive stars to produce ionizing photons. As a result, the principle ionizing sources vanish once the most massive stars disappear after the first few Myrs. In contrast, binary interactions and mergers can persist even after the most massive stars disappear, so binary models are capable of generating ionizing photons at late times.

\subsection{Ionizing Photons}
\label{section:ionizing_photons}
Starting from the SEDs as a function of age, $F_\lambda(t)$, we can compute the time evolution of the ionizing photon luminosity, $Q(t)$. Figures~\ref{fig:feh_m1_ionizing_photons} and \ref{fig:feh_p0_ionizing_photons} show $Q(t)$ from 1 to 30~Myr for a $10^6~\msun$ population at $\rm [Fe/H]=-1.0$ and $0.0$. The top, middle, and bottom panels show photons with wavelengths below 912~\AA, 504~\AA, and 228~\AA, respectively. We also show the BPASS binary and SB99+Geneva rotating model predictions for comparison. Note that the metallicities are not perfectly matched: for BPASS and SB99+Geneva, $Z=0.002$ and 0.001 for Figure~\ref{fig:feh_m1_ionizing_photons} and $Z=0.014$ and 0.02 for Figure~\ref{fig:feh_p0_ionizing_photons}. 

All of the models produce comparable hydrogen-ionizing photon output rates until $\sim3~\rm Myr$. Note that $Q$ is systematically offset in the SB99+Geneva model at very early times due to the lack of stars more massive than $300~\msun$. The discrepancies between different models become more pronounced at higher energies: for photons blueward of 228~\AA, the differences are substantial as early as $\rm 2~Myr$. This is because $Q$ becomes much more sensitive to the shape of the SED at progressively shorter wavelengths, where the predictions are sensitive to fast evolving stars and details of hot star SEDs. Overall, the SB99+Geneva models produce the softest radiation field (see also Figures~\ref{fig:feh_m1_sed} and \ref{fig:feh_p0_sed}). The FSPS+MIST models produce copious amounts of helium-ionizing photons after 2~Myr when the WR stars begin to appear, which is several orders of magnitude larger than the SB99+Geneva models. Again, the binary effects are responsible for the prolonged production of ionizing photons at very late times.

As described in Section~\ref{section:fsps}, assigning an appropriate spectrum to a WR star in the isochrone is a nontrivial task due to the high optical depth of the WR wind beyond the standard hydrostatic surface computed by stellar evolutionary codes. We tested the effect of modifying the WR temperature according to the \cite{Smith2002} weighting scheme, which is the default choice in SB99. Consistent with what \cite{Levesque2012} found, the difference is negligible for the hydrogen ionizing luminosity but becomes more pronounced for harder photons, which may explain the large discrepancy between the SB99+Geneva and FSPS+MIST predictions for helium-ionizing photons. Since this pertains to WR stars only, any variation due to the choice of WR temperatures disappears after $\sim5\textrm{--}6$~Myr. We proceed with our default WR $T_{\rm eff}$ assignment, but emphasize that the choice of this weighting scheme (or the lack thereof) will introduce some variation to the predicted output of the most energetic photons.

\subsection{Momentum Output}
Here we consider the momentum output from massive stars. Figures~\ref{fig:feh_m1_feedback} and \ref{fig:feh_p0_feedback} show the cumulative (top) and instantaneous (bottom) specific momentum injected in radiation, wind, and type II supernovae. Momentum injected via radiation is computed as follows:
\begin{equation}
\label{equation:rad_momentum}
\frac{dp_{\rm rad}}{dt} = (1+\tau_{\rm IR})\frac{L_{\rm bol}(t)}{c} \;\;,
\end{equation}
where $\tau_{\rm IR}$ is the infrared optical depth. In detail there is also a factor of $(1-\exp(-\tau_{\rm UV/optical}))$, but given the large optical depth of UV/optical photons, it reduces to a factor of order unity \citep[e.g.,][]{Agertz2013}. The $\tau_{\rm IR}$ term accounts for multiple absorption and re-radiation of the infrared photons in a very optically thick medium. Its preferred fiducial value is still under debate \citep[e.g.,][]{Murray2010, Andrews2011, Hopkins2011, Krumholz2012}. For simplicity, we assume $\tau_{\rm IR}=1$ in order to isolate the effect of SPS models alone, but empirical relations between $\tau_{\rm IR}$ and cluster mass (see e.g., Figure 3 in \citealt{Agertz2013}) suggest that $\tau_{\rm IR}$ can be as large as $\sim100$ for a $10^6~\msun$ cluster.

The momentum injection rate from a stellar wind is obtained using the mass loss rate and the wind speed:

\begin{equation}
\frac{dp_{\rm wind}}{dt} = \dot{M}v_{\rm wind}\;\;,
\end{equation}

The wind speed is estimated using the relation adopted in SB99 \citep{Leitherer1992}, which is comparable to the escape speed from the stellar photosphere:
\begin{equation}
\begin{split}
\log(v_{\rm wind})\;[{\rm km\;s^{-1}}] = 1.23-0.30\log(L/L_{\odot}) \\
+0.55\log(M/M_{\odot})+0.64\log(\teff\;[K]) \\
+0.13\log(Z/Z_\odot)\;\;.
\end{split}
\end{equation}

Finally, the momentum deposition per type II SN (SNII) event is 
\begin{equation}
\label{equation:sn_momentum}
p_{\rm SNII} = \frac{2E_{\rm SNII}}{v_{\rm SNII}}\;\;,
\end{equation}
where we assume that a typical SNII explosion outputs $10^{51}~\rm erg$ of kinetic energy (thermalization via shocks) and the ejecta are launched at $v_{\rm SNII}=3\times10^8~\rm cm\;s^{-1}$ \citep{Dale2015}. Moreover, we assume that only stars with initial masses between $8~\msun$ and $40~\msun$ terminate their lives as typical SNII and stars more massive than $40~\msun$ directly collapse to a black hole (\citealt{Fryer1999}, but see also e.g., \citealt{Sukhbold2014}). The corresponding momentum injection rate is

\begin{equation}
\label{equation:sn_mom_rate}
\frac{dp_{\rm SNII}}{dt} = p_{\rm SNII}\frac{dN_{\rm SNII}}{dt}\;\;,
\end{equation}
where the supernovae rate is obtained by integrating the IMF weight over the most massive star still alive in the previous and current time steps divided by the time interval.

We compare FSPS+MIST to SB99+Geneva, which is a popular choice for stellar population model in many galaxy simulations that attempt to include the effects of stellar feedback \citep[e.g.,][]{Hopkins2014, Agertz2015}. We use the SB99 outputs directly for the wind momentum and SNII rates, but we compute ourselves the radiation momentum using the bolometric luminosity reported by SB99 along with Equation~\ref{equation:rad_momentum}, and convert the SNII rates to the SNII momentum injection rate using Equations~\ref{equation:sn_momentum} and \ref{equation:sn_mom_rate}. The first SN explosion occurs at an earlier time in the SB99+Geneva model because their rate calculation assumes that all stars above $8~\msun$ explode as SNII. For the FSPS+MIST model, the onset of SNII is delayed at higher rotation rates due to the increased lifetimes (see also Figure~\ref{fig:ms_lifetimes}). Radiative momentum is generally dominant, though this may be an underestimate since we set the $\tau_{\rm IR}$ enhancement factor to unity. At a fixed rotation rate, the metal-poor population ($\rm [Fe/H]=-1.0$) outputs more radiative momentum compared to the metal-rich population, because metal-poor stars tend to be hotter and brighter.

\subsection{The Effects of Very Massive Stars}
\label{section:imf_upper_mass_limit}
The IMF dictates how frequently stars within a certain mass range occur in nature. Understanding the origin of and quantifying the high-mass IMF slope and the cutoff mass (and their dependence on environmental factors such as metallicity) are active areas of research \citep[e.g.,][]{Krumholz2011, Kroupa2013, Hopkins2013, Andrews2013, Narayanan2013, Dib2014, Weisz2015}. There are ongoing efforts to constrain these properties, but it is a difficult task due to their short lifetimes, rare numbers, and stochastic sampling effects. 

Here we investigate the importance of VMSs ($M>100~\msun$) in SPS modeling by comparing two otherwise identical models with different cutoff masses. The default option in SB99, a widely-used SPS package that is commonly paired with the Geneva stellar evolutionary models \citep{Ekstrom2012, Georgy2013, Yusof2013}, accounts for stars with initial masses up to $100~\msun$, though the underlying Geneva models are available up to $500~\msun$. The BPASS models are available with the upper mass limit set to either $100~\msun$ or $300~\msun$. In a recent observational work, \cite{Smith2016} found that VMSs, rather than less massive but very fast-rotating stars, are necessary to simultaneously explain the large ionizing flux and spectral emission line features observed in the nuclear star clusters in a blue compact dwarf galaxy NGC~5253. They stressed the need for SPS models to include stars more massive than $100~\msun$ in order to correctly predict the properties of young massive star clusters, including those residing in high-redshift star-forming galaxies.

\begin{figure}
\centering
\includegraphics[width=0.45\textwidth]{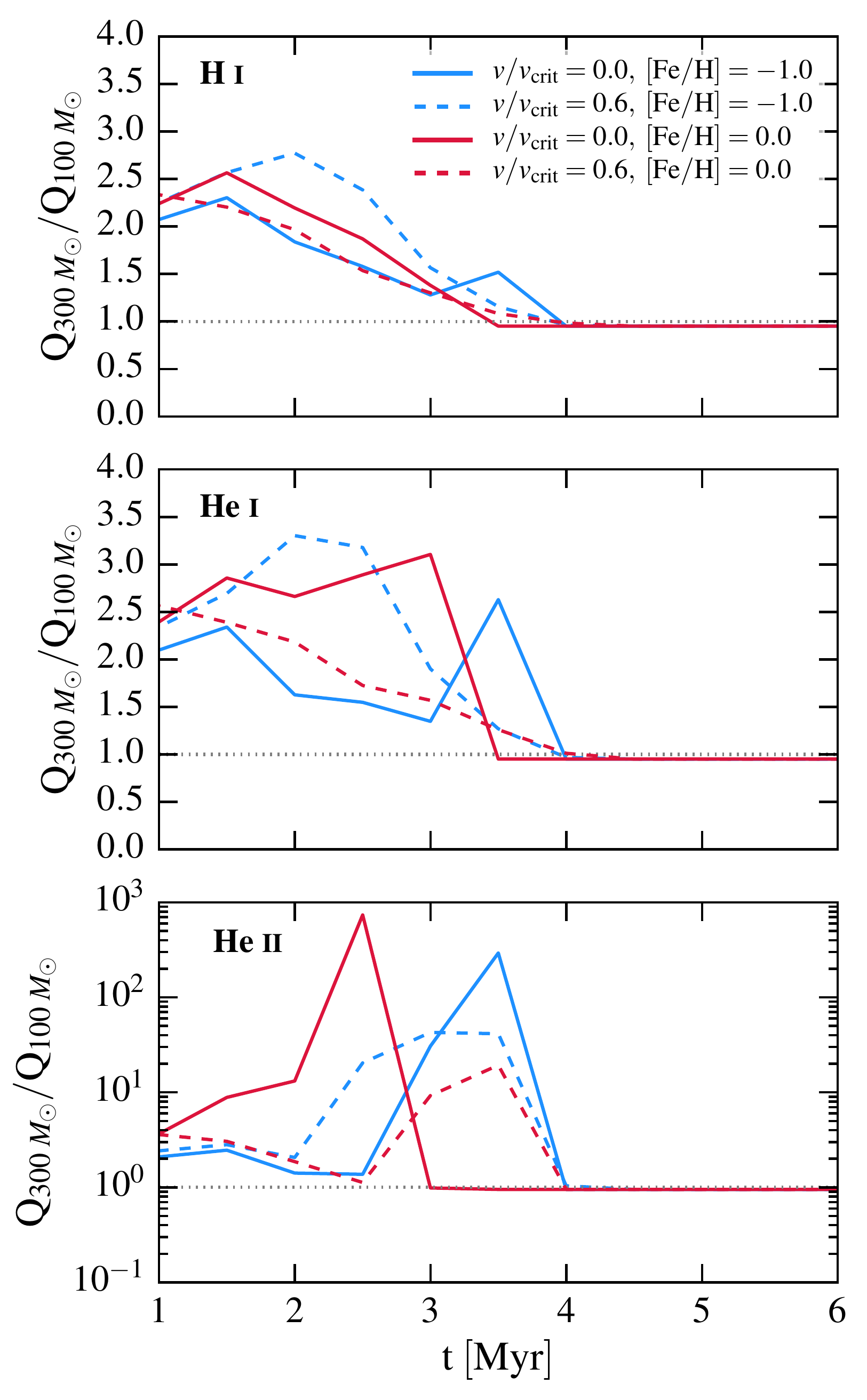} 
\caption{The ratio of the number of ionizing photons in the FSPS+MIST models with the IMF mass cutoff at $300~\msun$ and $100~\msun$ for two initial rotation rates and two metallicities. As expected, the ratio is essentially unity beyond $\sim4~\rm Myr$, by which time stars more massive than $100~\msun$ have disappeared. In detail, it is slightly less than unity due to the IMF weights. Top: photons capable of ionizing hydrogen. Middle: photons capable of singly ionizing helium. Bottom: photons capable of doubly ionizing helium. Note the vastly different $y$-axis range compared to the top two panels.} 
\label{fig:imf_300_100_ratio_ionizing}
\end{figure}

\begin{figure}
\centering
\includegraphics[width=0.45\textwidth]{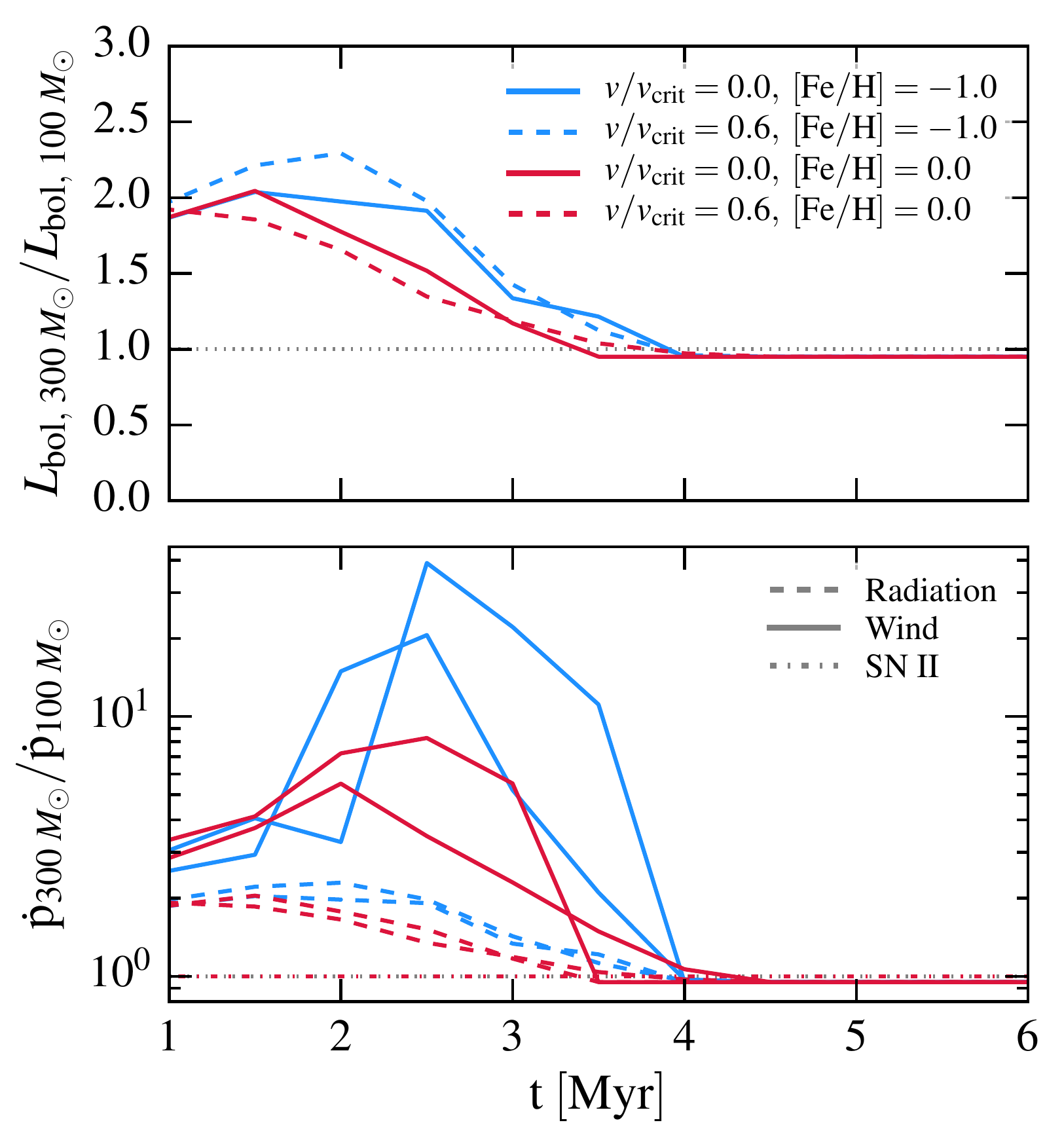} 
\caption{Top: the ratio of the bolometric luminosities in the FSPS+MIST models with the IMF mass cutoff at $300~\msun$ and $100~\msun$ for two initial rotation rates and two metallicities. Bottom: the same as the top panel except now showing the ratio of the instantaneous momentum injection.} 
\label{fig:imf_300_100_ratio_Lbol}
\end{figure}

Since the MIST model grid ranges from $0.1$ to $300~\msun$, we are able to examine the contribution from VMSs to the SED, more specifically the ionizing luminosity over time. Recall from Figure~\ref{fig:feh_zmet1_ionizing_photons_imf300} that the most massive stars contribute significantly to the flux blueward of 228~\AA{} at 1 and 2~Myr, and completely dominate at 3~Myr. Beyond 3~Myr, they contribute zero flux because they have all but disappeared.

Figure~\ref{fig:imf_300_100_ratio_ionizing} shows the ratio of ionizing luminosity emitted by stellar populations with IMF cutoffs at 300 and $100~\msun$. For simplicity, we show only the $v/v_{\rm crit}=0.0$ and 0.6 models. Similarly, Figure~\ref{fig:imf_300_100_ratio_Lbol} shows the ratio of the total bolometric luminosity and momentum output. As expected, the ratio is nearly unity beyond $\sim4~\rm Myr$, by which time stars more massive than $100~\msun$ have disappeared. In detail, it is slightly less than unity due to the IMF weights. Moreover, note that the ratio of the momentum injection from supernovae is unity at all times because we assume that only stars less massive than $40~\msun$ end their lives as SNII. These plots confirm what could be gleaned from Figure~\ref{fig:feh_zmet1_ionizing_photons_imf300}; the inclusion of VMSs can have an important effect on the integrated stellar population properties at very early times. Interestingly, this amounts to a factor of a few difference in some cases, e.g., hydrogen ionizing flux, bolometric luminosity, and radiative momentum output, whereas there can be an order of magnitude or more difference in the hardest EUV photon flux and wind momentum output. The dramatic difference in the helium ionizing photon luminosity can be understood by examining the contribution to the flux blueward of 228~\AA{} in Figure~\ref{fig:feh_zmet1_ionizing_photons_imf300}; the large peak between 2 and 4~Myr is directly linked to the inclusion (or exclusion) of massive stars completely stripped of their hydrogen envelope.

These results are in broad agreement with those reported by \cite{Smith2016}, who investigated the discrepancy in ages inferred from SED-fitting ($\sim1$~Myr; \citealt{Calzetti2015}) and WR spectroscopic features ($3\textrm{--}5$~Myr; \citealt{Turner2015}) of two nuclear star clusters in NGC~5253. They concluded that VMSs can explain the observed WR-like spectroscopic features, bringing the two age estimates into agreement ($1\textrm{--}2$~Myr) without needing to invoke older ages or extreme rotators. We separately computed the ionizing photon luminosity of a $v/v_{\rm crit}=0.4$, $\rm [Fe/H]=-0.50$ (the metallicity of NGC~5253 is $35\%$ of the solar value assuming the \citealt{Asplund2009} solar oxygen abundance; \citealt{MonrealIbero2012}) stellar population both with and without stars more massive than $100~\msun$. Following \cite{Smith2016}, we assumed a total stellar mass of $3.3\times10^5~\msun$ for the two clusters and compared the ionizing flux at 2 and 4~Myr to the observed value of $2.2\times10^{52}~\rm s^{-1}$ for the central 5~pc. When we set the IMF upper limit at $100~\msun$, we obtain $1.7\times10^{52}~\rm s^{-1}$ and $1.0\times10^{52}~\rm s^{-1}$ at 2 and 4~Myr, respectively. Importantly, the maximum value of the ionizing flux attained by a population excluding VMSs is only $\sim1.9\times10^{52}~\rm s^{-1}$. We confirm the \cite{Smith2016} result that models that do not account for VMSs underpredict the ionizing flux. However, when we extend the upper mass limit to $300~\msun$, we obtain $4.0\times10^{52}~\rm s^{-1}$ (maximum) and $0.9\times10^{52}~\rm s^{-1}$ at 2 and 4~Myr. The conclusions do not change qualitatively when we instead consider a non-rotating stellar population. However, some rotational mixing resulting from a moderate rotation is necessary in order to match the observed factor of $\sim3$ enhancement in the nebular nitrogen abundance reported by \cite{Smith2016}. Both the excess and the enrichment timescale for nitrogen in the MIST rotating model are in agreement with the \cite{Koehler2015} evolutionary model, which was demonstrated in \cite{Smith2016} to be capable of producing the level of enrichment observed in the clusters.

The main conclusion here is that for most integrated quantities of interest, such as the hydrogen ionizing luminosity, increasing the IMF upper mass cutoff to $300~\msun$ results in a factor of a few increase at very early times ($t\lesssim3$~Myr; see also \citealt{Stanway2016}). This difference may still have important observational consequences, as discussed in \cite{Smith2016}. For other quantities, such as the production of the most extreme EUV photons, the difference can be as large as several orders of magnitude, though these quantities are more sensitive to the underlying models of VMSs, which still have large uncertainties.

\subsection{The Effects of Metallicity}
Although we focus on $\rm [Fe/H]=-1.0$ and 0.0 stellar populations in this work to enable a roughly equal comparison across different models,\footnote{As a reminder, we compare with SB99+Geneva models at $Z=0.002$ and $Z=0.014$ and BPASS models at $Z=0.001$, and $Z=0.020$.} it is important to explore the effects of metallicity over a broader range. Metal-poor populations are particularly worth investigating since the effect of rotation on stellar population properties becomes more significant at lower metallicities. Moreover, as we will discuss in Section~\ref{section:discussion}, high-redshift galaxies believed to be the principle sources of ionizing photons have stellar metallicities of $\log(Z/Z_{\odot})\lesssim-2$ at $z=6$ \citep[e.g.,][]{Ma2016a}. 
 
\begin{figure}
\centering
\includegraphics[width=0.45\textwidth]{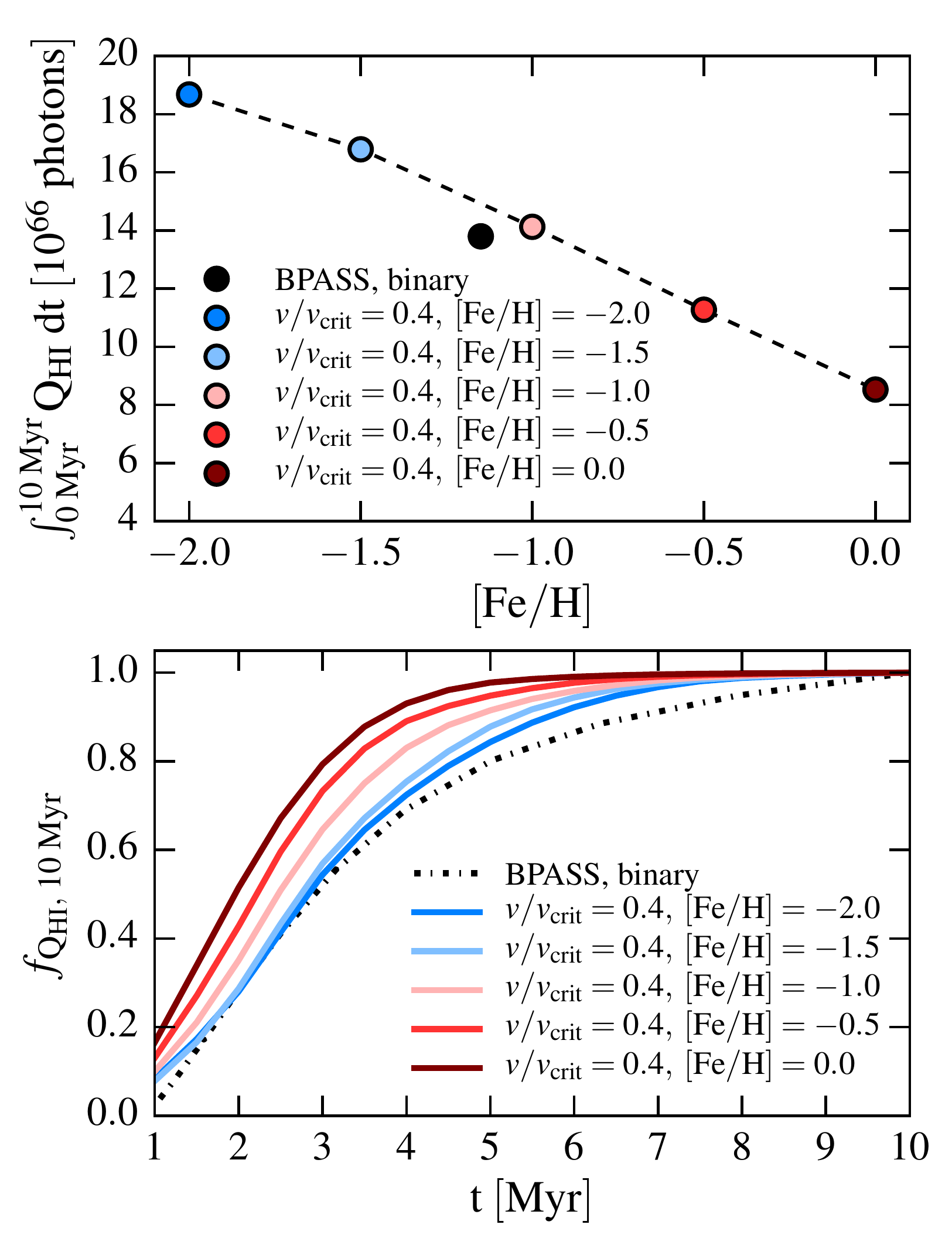} 
\caption{Top: the total number of hydrogen ionizing photons produced by 10~Myr. The different colors correspond to rotating stellar populations at different metallicities. As expected, the ionizing photon production is more efficient in low metallicity environments. For comparison, the equivalent point for the lowest metallicity BPASS binary model available ($Z=0.001$) is shown in black. For the BPASS model, integrating out to 30~Myr instead of 10~Myr makes a $\sim10\%$ difference in the total number of photons. Bottom: the fraction of total hydrogen-ionizing photons emitted by 10~Myr for the same set of models. The decrease in ionizing photon production with time is more gradual at lower metallicities.} 
\label{fig:feh_m15_ionizing_photons}
\end{figure}

Figure~\ref{fig:feh_m15_ionizing_photons} illustrates the effect of metallicity on the ionizing photon production. In the top panel, we show the total number of hydrogen ionizing photons produced by 10~Myr. The different colors correspond to rotating stellar populations at different metallicities. As expected, the ionizing photon production is more efficient in low metallicity environments. For comparison, the equivalent point for the lowest metallicity BPASS binary model available ($ Z=0.001$) is shown in black. We note that for the BPASS model, integrating out to 30~Myr instead of 10~Myr makes a $\sim10\%$ difference in the total number of photons. The bottom panel shows the fraction of total hydrogen-ionizing photons emitted by 10~Myr for the same set of models. The decrease in ionizing photon production with time is more gradual at lower metallicities. We revisit these points in Section~\ref{section:discussion}. 

\begin{figure}
\centering
\includegraphics[width=0.43\textwidth]{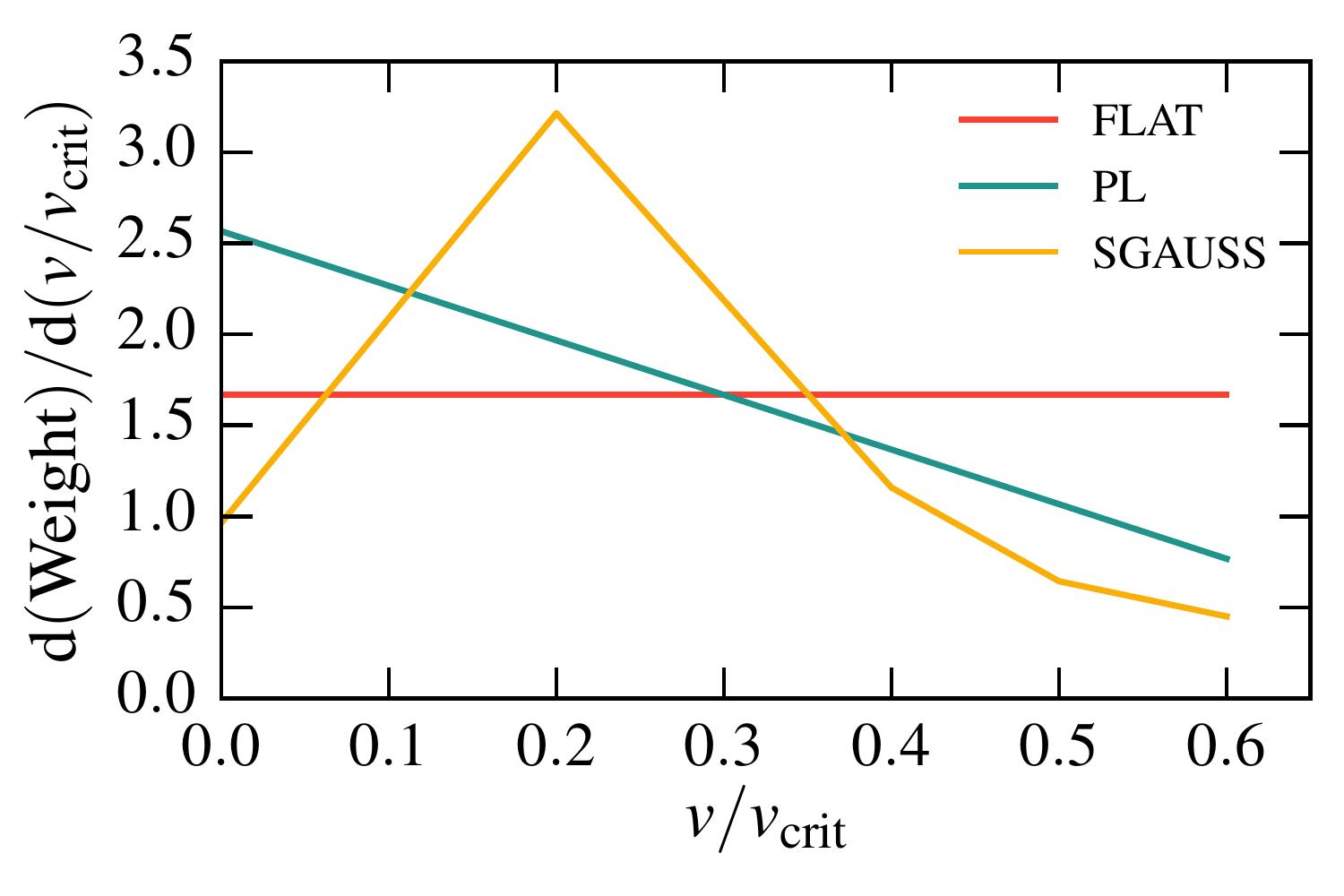} 
\caption{A schematic diagram showing the three types of rotational probably distribution functions explored in this work. ``FLAT'' represents a simple case of a flat distribution. ``PL'' is a simple power law that decreases with the rotation rate. ``SGAUSS'' is meant to approximate a skewed Gaussian distribution centered at $v/v_{\rm crit}=0.2$.}
\label{fig:vel_schematic}
\end{figure}

\begin{figure*}
\centering
\includegraphics[width=0.85\textwidth]{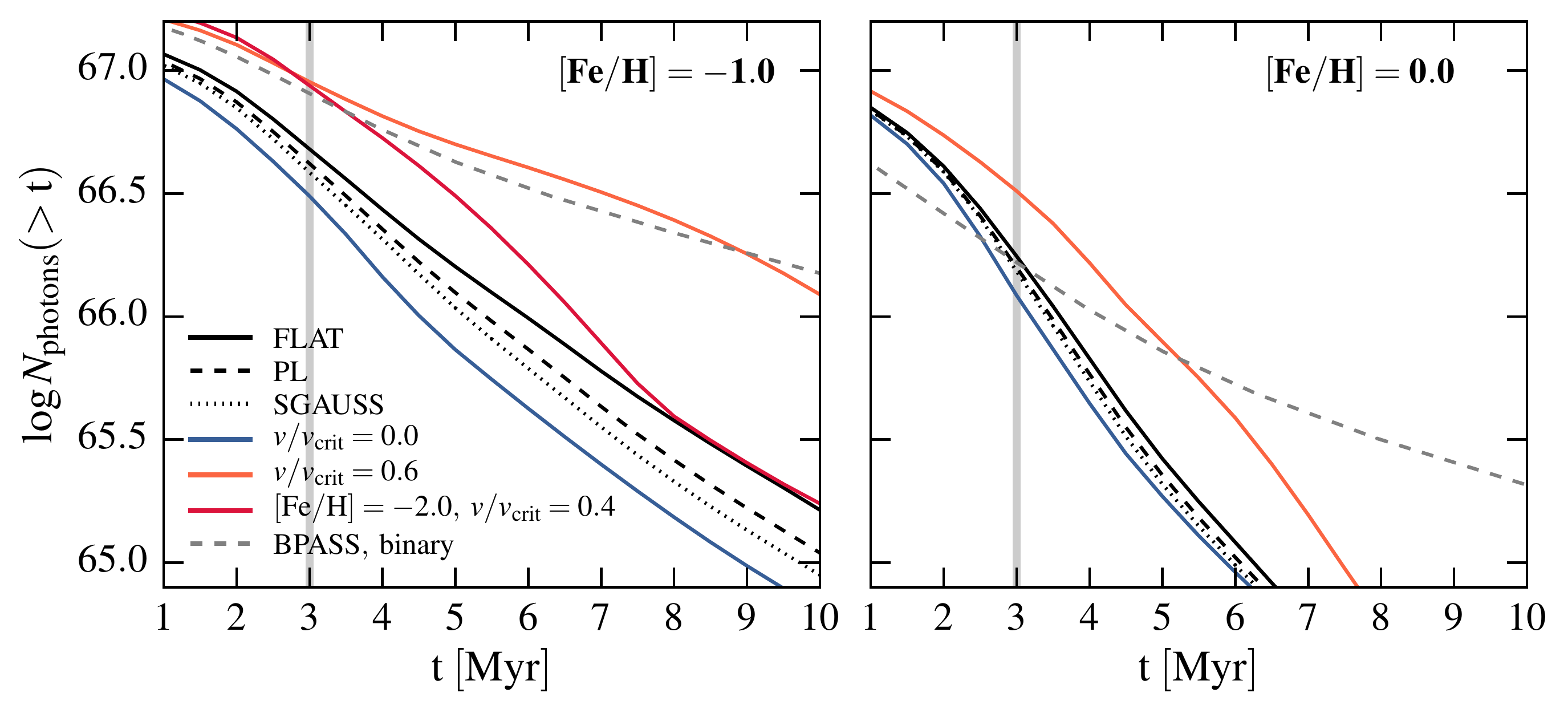} 
\caption{The total number of hydrogen-ionizing photons emitted between time $t$ and 30~Myr. Blue and orange curves correspond to the $v/v_{\rm crit}=0.0$ and 0.6 models at $\rm [Fe/H]=-1.0$ (left) and 0.0 (right). ``FLAT'' represents a simple case of a flat distribution, ``PL'' corresponds to a power law that decreases with the rotation rate, and ``SGAUSS'' approximates a skewed Gaussian distribution centered at $v/v_{\rm crit}=0.2$. The BPASS model assumes an IMF cutoff of $300~\msun$ for $Z=0.001$ (left) and $Z=0.014$(right). In the left panel, we also show the $v/v_{\rm crit}=0.4$ model at $\rm [Fe/H]=-2.0$ for comparison. The gray vertical line marks roughly when GMCs begin to disperse \citep{Ma2015}, yielding order unity escape fraction of ionizing photons.}
\label{fig:PDF_photons}
\end{figure*}

\subsection{The Effects of a Rotation Distribution}
\label{section:rotation_dist}
Observations of massive stars in Galactic and local group clusters show a distribution of rotation rates \citep[e.g.,][]{Penny1996, Huang2010, RamirezAgudelo2013}, which may hold important clues for shedding light on their formation processes \citep[e.g.,][]{Zinnecker2007}. A distribution of rotation rates is interesting in the context of this work for the simple reason that it would introduce {\it a range of lifetimes, and therefore a range of MSTO masses at a fixed age}. For example, by the time the dense GMC is sufficiently dispersed, allowing for ionizing photons to escape, the most massive slowly rotating stars will have already died, perhaps as SNe. However, the most massive rapidly rotating stars will still be present and producing copious quantities of ionizing photons.

For this work, recall that we computed MIST isochrones at five different rotation rates---$v/v_{\rm crit}=0.0, 0.2, 0.4, 0.5$, and $0.6$---for each metallicity. From these, we create three composite populations by combining SSPs with different $v/v_{\rm crit}$ according to the probability distribution functions (PDFs) shown in Figure~\ref{fig:vel_schematic}. ``FLAT'' represents a simple case of a flat distribution. ``PL'' is a simple power law that decreases with the rotation rate. ``SGAUSS'' approximates a skewed Gaussian distribution centered at $v/v_{\rm crit}=0.2$, which is supposed to mimic the observed distribution of O-type stars in the 30 Doradus region of the Large Magellanic Cloud \citep{RamirezAgudelo2013}. We simplify our approach by assuming that the rotation distribution is mass-independent over the mass range of interest ($M>20~\msun$). This is an acceptable assumption for two reasons. First, the rotation rate measurements of stars more massive than $M\sim40~\msun$ are scarce---for instance, $<20\%$ of the \citealt{RamirezAgudelo2013} sample was estimated to have masses above $40~\msun$---so their rotation rates, let alone their mass-dependence, are still relatively uncertain. Second, the observed rotation rates do not necessarily represent the rotation rates at birth: processes such as stellar winds, magnetic braking, and binary interactions modify the surface rotation rates, and the angular momentum is redistributed in the interior through mechanisms that are not fully understood in stars of all masses \citep[e.g.,][]{Wolff2004, Cantiello2010, Gallet2013, Fuller2015}. However, studying a select sample of young MS stars somewhat alleviates this issue (for example, \citealt{Huang2010} uses a $\log g$ criterion). Further complications include macroturbulence \citep[e.g.,][]{Penny2009} and $\sin i$ projection effects, which make the task of inferring the initial conditions and the subsequent evolution much more complex. We attempt to mitigate some of these problems by exploring a range of rotation PDFs. Nevertheless, given these complexities, we emphasize that this is an attempt to explore the broad effects of a rotation distribution on observed quantities. 

In Figure~\ref{fig:PDF_photons}, we plot the total number of hydrogen-ionizing photons emitted from time $t$ to 30~Myr for $\rm [Fe/H]=-1.0$ (left) and $0.0$ (right). The blue and orange curves correspond to $v/v_{\rm crit}=0.0$ and 0.6 single-rotation models and the gray dashed line represents the BPASS binary model, which assumes an IMF cutoff of $300~\msun$ for $Z=0.001$ (left) and $Z=0.014$ (right). In the left panel, we additionally show the $v/v_{\rm crit}=0.4$ model at $\rm [Fe/H]=-2.0$ for comparison. The gray vertical line marks roughly when GMCs begin to disperse, yielding order unity escape fraction of ionizing photons out of GMCs \citep[e.g.,][]{Ma2015}. 

A composite population with a flat rotational distribution (``FLAT'')---which has the largest contribution from $v/v_{\rm crit}=0.6$---still has three times fewer photons in its budget compared to the fastest-rotating and binary scenarios by the time GMCs begin to disperse. The very low-metallicity case ($\rm [Fe/H]=-2.0$), where the rotation effects are expected to be large and therefore the fall-off in the production rate of ionizing photons is more gradual in time (Figure~\ref{fig:feh_m15_ionizing_photons}), has a comparable ionizing photon budget compared to the fast-rotating and binary scenarios by 3~Myr. However the number of available photons falls by an order of magnitude below that of the binary model by 8~Myr.

Interestingly, a very rapidly rotating population with $v/v_{\rm crit}=0.6$ is able to mimic the time-dependence of ionizing luminosity in a binary population. However, observations of single O-type stars in the LMC \citep{RamirezAgudelo2013} effectively rule out populations of very fast-rotating massive stars, at least in the local Universe. It is also interesting to note that the ionizing luminosity in a moderately rotating stellar population at very low metallicity can rival that of the binary model and rapidly rotating model, though the production rate also falls off steeply beyond $\sim5\textrm{--}6$~Myr.

\vspace{0.5cm}
\section{Discussion}
\label{section:discussion}
In this section we discuss the implications of the rotating, massive star models in two main contexts: cosmic reionization and the interpretation of high-redshift star-forming galaxies.

In our current understanding of the high-redshift universe, cosmic reionization was largely driven by high energy photons leaking out from star-forming galaxies \citep[e.g.,][]{Madau1999, Haardt2012}. The escape fraction of these ionizing photons, $f_{\rm esc}$, required to explain the observed ionization state of the $z=6$ intergalactic medium exceeds $10\%$ \citep[e.g.,][]{Finkelstein2012, Robertson2013, Gnedin2014}, well above the values inferred from their lower redshift analogs \citep[e.g.,][]{Iwata2009, Boutsia2011, Siana2010, Leitet2013}. Galaxy simulations have produced a wide range of predicted $f_{\rm esc}$ \citep[e.g.,][]{Gnedin2008, Razoumov2010, Wise2014}. In particular, a recent suite of simulations from the Feedback in Realistic Environment (FIRE) project produced a time-averaged $f_{\rm esc}$ of only $\lesssim5\%$ \citep{Ma2015}. Since the production of ionizing photons is dominated by young massive stellar populations embedded in optically-thick natal environments, $f_{\rm esc}$ from these simulations depends sensitively on the complex connection between the properties of the underlying stellar population model\footnote{The FIRE simulations use SB99 with the ``Padova+Asymptotic Giant Branch'' option \citep{Bressan1993, Fagotto1994a, Fagotto1994b, Girardi2000}.} and the mechanisms that disrupt the GMCs \citep[e.g.,][]{Murray2010}. In their follow-up work, \cite{Ma2016b} found that replacing the underlying stellar population model with one that includes binary effects (BPASS) in their post-processing radiative transfer calculations can boost $f_{\rm esc}$ significantly due to its ability to sustain ionizing photon production at late times, well after the disruption of the GMCs. As emphasized by the authors, the key difference between the fiducial single star models and the binary models is the time-dependence of the ionizing photon production. Other mechanisms have also been proposed to increase the escape fraction of ionizing radiation, including runaway massive stars \citep{Conroy2012, Kimm2014}.

In Section~\ref{section:rotation_dist}, we explored the same concept by considering low metallicities and velocity distributions as means to prolong the ionizing photon production. Together, Figures~\ref{fig:feh_m15_ionizing_photons} and \ref{fig:PDF_photons} demonstrate that stellar populations in sufficiently low-metallicity environments (e.g., high redshift) require only moderate rotation rates in order to produce significant ionizing photons. Moreover, at the time of order unity escape fraction ($\sim3$~Myr), the ionizing photon budget can rival that of a binary population or a fast-rotating but slightly more metal-rich population. However, it is very challenging to {\t prolong} the emission of ionizing photons by an amount required to reproduce the relatively large values of $f_{\rm esc}$ implied by the cosmic reionization models \citep{Ma2016b}, and we conclude that binary interactions may indeed play a critical role. Detailed numerical simulations using these low-metallicity rotating stellar population models are required to address this point more quantitatively. 

Rotating massive stars have also been considered in the context of high-redshift star-forming galaxies. Recently, \cite{Steidel2014} have found that the locus of $z\sim2\textrm{--}3$ star-forming galaxies is offset relative to their $z=0$ counterparts in the BPT diagram \citep{Baldwin1981}, an empirical diagnostic for probing the physical conditions and ionizing sources of nebular gas. The authors concluded that this shift could be explained by a harder stellar ionizing radiation field ($T_{\rm blackbody}=50000\textrm{--}60000$~K), higher ionization parameter (the ratio of the number densities of hydrogen ionizing photons to hydrogen atoms in a \ionn{H}{ii} region), and nitrogen-enhanced nebular gas, all of which could be explained by rapid rotation or binaries. In subsequent work, \cite{Steidel2016} found that the BPASS binary models are able to generate sufficiently hard ionizing radiation fields to reproduce the various observed line ratios, while the Geneva+SB99 cannot self-consistently explain the observations at any gas-phase metallicity or ionization parameter. Figure~\ref{fig:feh_m1_sed} shows that the FSPS+MIST models predict EUV flux that falls right in between the two models considered in \cite{Steidel2016}. A more detailed comparison between the emission line predictions for the FSPS+MIST, Geneva+SB99, and BPASS models will be carried out in Byler et al. (2017, submitted).

\section{Conclusions}
\label{section:conclusions}
In this work we explored the integrated properties of massive, rotating single-star stellar populations in both metal-rich and metal-poor environments. Our main conclusions are as follows:

\begin{enumerate}
\item We confirm that rotation leads to longer MS lifetimes and brighter and hotter stars. We also find that the magnitude of the effects is non-linear with the rotation rate. In particular, the $v/v_{\rm crit}=0.5$ and 0.6 models show a significant enhancement in the ionizing luminosity due to the blueward evolution of massive stars in the HR diagram at ages greater than 3~Myr. A comparison of the predicted MS lifetimes for the Geneva \citep{Ekstrom2012, Georgy2013, Yusof2013} and MIST \citep{Dotter2016, Choi2016} stellar evolutionary tracks demonstrates that the MS lifetime boost at a fixed mass is larger in the Geneva models compared to that in the MIST models, which suggests that rotational mixing may be more efficient in the former.

\item Rotation effects become more significant at lower metallicities as the star becomes more compact and angular momentum loss due to winds becomes less important. We found that the $\rm [Fe/H] = -2.0$ population produces twice as many photons as the $\rm [Fe/H] = 0.0$ population by 10~Myr. From comparisons among models ranging from $\rm [Fe/H] = 0.0$ to $-2.0$, we conclude that rotation leads to a more gradual decline in ionizing luminosity with time at low metallicities. As demonstrated in \cite{Ma2016b}, this time-dependence has interesting ramifications for the escape fraction of ionizing photons, which in turn could impact high redshift, metal-poor galaxies at the time of cosmic reionization.

\item The contribution from VMSs ($\geq100~\msun$) is significant for a short period of time ($t\lesssim4$~Myr), increasing the ionizing luminosity by a factor of a few (\ionn{H}{i}) to a few orders of magnitude (\ionn{He}{ii}). Although the cumulative effect is small in most cases, taking VMSs into account in SPS modeling may be important for understanding a number of recent observations, including the ages and ionizing photon budget in nuclear star clusters \citep{Smith2016}. 

\item We explored composite stellar populations with three different rotation PDFs to investigate whether a small number of fast rotators can lead to a significant boost in the ionizing photon luminosity at late times ($\sim10$~Myr) compared to a non-rotating population. We find that the effect is small, amounting to a factor of two or smaller for a skewed distribution peaking at low $\rm v/v_{\rm crit}$ and a factor of $\sim3$ for the extreme case of a flat PDF. Numerical simulations are required to assess whether or not this effect is important for the disruption of GMCs and/or the escape of ionizing radiation from galaxies.

\end{enumerate}

\acknowledgments{}
We thank J.J. Eldridge, Mason Ng, and Georgie Taylor for sharing with us their unpublished WM-Basic models. We thank Matteo Cantiello, Aaron Dotter, and Daniel Weisz for useful comments and suggestions that have improved the manuscript. We also thank Bill Paxton and the rest of the MESA community who have made this project possible. C.C. acknowledges support from NASA grant NNX13AI46G, NSF grant AST- 1313280, and the Packard Foundation.

\bibliographystyle{apj}
\bibliography{bibtex.bib}

\begin{thebibliography}{}
\expandafter\ifx\csname natexlab\endcsname\relax\def\natexlab#1{#1}\fi

\bibitem[{{Agertz} \& {Kravtsov}(2015)}]{Agertz2015}
{Agertz}, O., \& {Kravtsov}, A.~V. 2015, \apj, 804, 18

\bibitem[{{Agertz} {et~al.}(2013){Agertz}, {Kravtsov}, {Leitner}, \&
  {Gnedin}}]{Agertz2013}
{Agertz}, O., {Kravtsov}, A.~V., {Leitner}, S.~N., \& {Gnedin}, N.~Y. 2013,
  \apj, 770, 25

\bibitem[{{Agertz} {et~al.}(2009){Agertz}, {Lake}, {Teyssier}, {Moore},
  {Mayer}, \& {Romeo}}]{Agertz2009}
{Agertz}, O., {Lake}, G., {Teyssier}, R., {et~al.} 2009, \mnras, 392, 294

\bibitem[{{Andrews} \& {Thompson}(2011)}]{Andrews2011}
{Andrews}, B.~H., \& {Thompson}, T.~A. 2011, \apj, 727, 97

\bibitem[{{Andrews} {et~al.}(2013){Andrews}, {Calzetti}, {Chandar}, {Lee},
  {Elmegreen}, {Kennicutt}, {Whitmore}, {Kissel}, {da Silva}, {Krumholz},
  {O'Connell}, {Dopita}, {Frogel}, \& {Kim}}]{Andrews2013}
{Andrews}, J.~E., {Calzetti}, D., {Chandar}, R., {et~al.} 2013, \apj, 767, 51

\bibitem[{{Asplund} {et~al.}(2009){Asplund}, {Grevesse}, {Sauval}, \&
  {Scott}}]{Asplund2009}
{Asplund}, M., {Grevesse}, N., {Sauval}, A.~J., \& {Scott}, P. 2009, \araa, 47,
  481

\bibitem[{{Baldwin} {et~al.}(1981){Baldwin}, {Phillips}, \&
  {Terlevich}}]{Baldwin1981}
{Baldwin}, J.~A., {Phillips}, M.~M., \& {Terlevich}, R. 1981, \pasp, 93, 5

\bibitem[{{Barbaro} \& {Bertelli}(1977)}]{Barbaro1977}
{Barbaro}, C., \& {Bertelli}, C. 1977, \aap, 54, 243

\bibitem[{{Boutsia} {et~al.}(2011){Boutsia}, {Grazian}, {Giallongo}, {Fontana},
  {Pentericci}, {Castellano}, {Zamorani}, {Mignoli}, {Vanzella}, {Fiore},
  {Lilly}, {Gallozzi}, {Testa}, {Paris}, \& {Santini}}]{Boutsia2011}
{Boutsia}, K., {Grazian}, A., {Giallongo}, E., {et~al.} 2011, \apj, 736, 41

\bibitem[{{Bressan} {et~al.}(1993){Bressan}, {Fagotto}, {Bertelli}, \&
  {Chiosi}}]{Bressan1993}
{Bressan}, A., {Fagotto}, F., {Bertelli}, G., \& {Chiosi}, C. 1993, \aaps, 100,
  647

\bibitem[{{Brott} {et~al.}(2011){Brott}, {de Mink}, {Cantiello}, {Langer}, {de
  Koter}, {Evans}, {Hunter}, {Trundle}, \& {Vink}}]{Brott2011}
{Brott}, I., {de Mink}, S.~E., {Cantiello}, M., {et~al.} 2011, \aap, 530, A115

\bibitem[{{Calzetti} {et~al.}(2015){Calzetti}, {Johnson}, {Adamo}, {Gallagher},
  {Andrews}, {Smith}, {Clayton}, {Lee}, {Sabbi}, {Ubeda}, {Kim}, {Ryon},
  {Thilker}, {Bright}, {Zackrisson}, {Kennicutt}, {de Mink}, {Whitmore},
  {Aloisi}, {Chandar}, {Cignoni}, {Cook}, {Dale}, {Elmegreen}, {Elmegreen},
  {Evans}, {Fumagalli}, {Gouliermis}, {Grasha}, {Grebel}, {Krumholz},
  {Walterbos}, {Wofford}, {Brown}, {Christian}, {Dobbs}, {Herrero}, {Kahre},
  {Messa}, {Nair}, {Nota}, {{\"O}stlin}, {Pellerin}, {Sacchi}, {Schaerer}, \&
  {Tosi}}]{Calzetti2015}
{Calzetti}, D., {Johnson}, K.~E., {Adamo}, A., {et~al.} 2015, \apj, 811, 75

\bibitem[{{Cantiello} \& {Langer}(2010)}]{Cantiello2010}
{Cantiello}, M., \& {Langer}, N. 2010, \aap, 521, A9

\bibitem[{{Castor} {et~al.}(1975){Castor}, {Abbott}, \& {Klein}}]{Castor1975}
{Castor}, J.~I., {Abbott}, D.~C., \& {Klein}, R.~I. 1975, \apj, 195, 157

\bibitem[{{Cervi{\~n}o} {et~al.}(2001){Cervi{\~n}o}, {G{\'o}mez-Flechoso},
  {Castander}, {Schaerer}, {Moll{\'a}}, {Kn{\"o}dlseder}, \&
  {Luridiana}}]{Cervino2001}
{Cervi{\~n}o}, M., {G{\'o}mez-Flechoso}, M.~A., {Castander}, F.~J., {et~al.}
  2001, \aap, 376, 422

\bibitem[{{Cervi{\~n}o} {et~al.}(2000){Cervi{\~n}o}, {Luridiana}, \&
  {Castander}}]{Cervino2000}
{Cervi{\~n}o}, M., {Luridiana}, V., \& {Castander}, F.~J. 2000, \aap, 360, L5

\bibitem[{{Cervi{\~n}o} {et~al.}(2002){Cervi{\~n}o}, {Valls-Gabaud},
  {Luridiana}, \& {Mas-Hesse}}]{Cervino2002}
{Cervi{\~n}o}, M., {Valls-Gabaud}, D., {Luridiana}, V., \& {Mas-Hesse}, J.~M.
  2002, \aap, 381, 51

\bibitem[{{Chini} {et~al.}(2012){Chini}, {Hoffmeister}, {Nasseri}, {Stahl}, \&
  {Zinnecker}}]{Chini2012}
{Chini}, R., {Hoffmeister}, V.~H., {Nasseri}, A., {Stahl}, O., \& {Zinnecker},
  H. 2012, \mnras, 424, 1925

\bibitem[{{Choi} {et~al.}(2016){Choi}, {Dotter}, {Conroy}, {Cantiello},
  {Paxton}, \& {Johnson}}]{Choi2016}
{Choi}, J., {Dotter}, A., {Conroy}, C., {et~al.} 2016, \apj, 823, 102

\bibitem[{{Conroy} \& {Gunn}(2010)}]{Conroy2010}
{Conroy}, C., \& {Gunn}, J.~E. 2010, \apj, 712, 833

\bibitem[{{Conroy} {et~al.}(2009){Conroy}, {Gunn}, \& {White}}]{Conroy2009}
{Conroy}, C., {Gunn}, J.~E., \& {White}, M. 2009, \apj, 699, 486

\bibitem[{{Conroy} \& {Kratter}(2012)}]{Conroy2012}
{Conroy}, C., \& {Kratter}, K.~M. 2012, \apj, 755, 123

\bibitem[{{Creasey} {et~al.}(2013){Creasey}, {Theuns}, \&
  {Bower}}]{Creasey2013}
{Creasey}, P., {Theuns}, T., \& {Bower}, R.~G. 2013, \mnras, 429, 1922

\bibitem[{{Crowther} {et~al.}(2016){Crowther}, {Caballero-Nieves}, {Bostroem},
  {Ma{\'{\i}}z Apell{\'a}niz}, {Schneider}, {Walborn}, {Angus}, {Brott},
  {Bonanos}, {de Koter}, {de Mink}, {Evans}, {Gr{\"a}fener}, {Herrero},
  {Howarth}, {Langer}, {Lennon}, {Puls}, {Sana}, \& {Vink}}]{Crowther2016}
{Crowther}, P.~A., {Caballero-Nieves}, S.~M., {Bostroem}, K.~A., {et~al.} 2016,
  \mnras, 458, 624

\bibitem[{{da Silva} {et~al.}(2012){da Silva}, {Fumagalli}, \&
  {Krumholz}}]{daSilva2012}
{da Silva}, R.~L., {Fumagalli}, M., \& {Krumholz}, M. 2012, \apj, 745, 145

\bibitem[{{da Silva} {et~al.}(2014){da Silva}, {Fumagalli}, \&
  {Krumholz}}]{daSilva2014}
{da Silva}, R.~L., {Fumagalli}, M., \& {Krumholz}, M.~R. 2014, \mnras, 444,
  3275

\bibitem[{{Dale}(2015)}]{Dale2015}
{Dale}, J.~E. 2015, New Astronomy Review, 68, 1

\bibitem[{{de Avillez} \& {Breitschwerdt}(2004)}]{deAvillez2004}
{de Avillez}, M.~A., \& {Breitschwerdt}, D. 2004, \aap, 425, 899

\bibitem[{{de Jager} {et~al.}(1988){de Jager}, {Nieuwenhuijzen}, \& {van der
  Hucht}}]{deJager1988}
{de Jager}, C., {Nieuwenhuijzen}, H., \& {van der Hucht}, K.~A. 1988, \aaps,
  72, 259

\bibitem[{{de Mink} {et~al.}(2014){de Mink}, {Sana}, {Langer}, {Izzard}, \&
  {Schneider}}]{deMink2014}
{de Mink}, S.~E., {Sana}, H., {Langer}, N., {Izzard}, R.~G., \& {Schneider},
  F.~R.~N. 2014, \apj, 782, 7

\bibitem[{{Dekel} \& {Silk}(1986)}]{Dekel1986}
{Dekel}, A., \& {Silk}, J. 1986, \apj, 303, 39

\bibitem[{{Dib}(2014)}]{Dib2014}
{Dib}, S. 2014, \mnras, 444, 1957

\bibitem[{{Dotter}(2016)}]{Dotter2016}
{Dotter}, A. 2016, \apjs, 222, 8

\bibitem[{{Dove} \& {Shull}(1994)}]{Dove1994}
{Dove}, J.~B., \& {Shull}, J.~M. 1994, \apj, 430, 222

\bibitem[{{Eggenberger} {et~al.}(2008){Eggenberger}, {Meynet}, {Maeder},
  {Hirschi}, {Charbonnel}, {Talon}, \& {Ekstr{\"o}m}}]{Eggenberger2008}
{Eggenberger}, P., {Meynet}, G., {Maeder}, A., {et~al.} 2008, \apss, 316, 43

\bibitem[{{Ekstr{\"o}m} {et~al.}(2012){Ekstr{\"o}m}, {Georgy}, {Eggenberger},
  {Meynet}, {Mowlavi}, {Wyttenbach}, {Granada}, {Decressin}, {Hirschi},
  {Frischknecht}, {Charbonnel}, \& {Maeder}}]{Ekstrom2012}
{Ekstr{\"o}m}, S., {Georgy}, C., {Eggenberger}, P., {et~al.} 2012, \aap, 537,
  A146

\bibitem[{{Eldridge}(2012)}]{Eldridge2012}
{Eldridge}, J.~J. 2012, \mnras, 422, 794

\bibitem[{{Eldridge} \& {Stanway}(2009)}]{Eldridge2009}
{Eldridge}, J.~J., \& {Stanway}, E.~R. 2009, \mnras, 400, 1019

\bibitem[{{Endal} \& {Sofia}(1978)}]{Endal1978}
{Endal}, A.~S., \& {Sofia}, S. 1978, \apj, 220, 279

\bibitem[{{Evans} {et~al.}(2009){Evans}, {Dunham}, {J{\o}rgensen}, {Enoch},
  {Mer{\'{\i}}n}, {van Dishoeck}, {Alcal{\'a}}, {Myers}, {Stapelfeldt},
  {Huard}, {Allen}, {Harvey}, {van Kempen}, {Blake}, {Koerner}, {Mundy},
  {Padgett}, \& {Sargent}}]{Evans2009}
{Evans}, II, N.~J., {Dunham}, M.~M., {J{\o}rgensen}, J.~K., {et~al.} 2009,
  \apjs, 181, 321

\bibitem[{{Fagotto} {et~al.}(1994{\natexlab{a}}){Fagotto}, {Bressan},
  {Bertelli}, \& {Chiosi}}]{Fagotto1994a}
{Fagotto}, F., {Bressan}, A., {Bertelli}, G., \& {Chiosi}, C.
  1994{\natexlab{a}}, \aaps, 104

\bibitem[{{Fagotto} {et~al.}(1994{\natexlab{b}}){Fagotto}, {Bressan},
  {Bertelli}, \& {Chiosi}}]{Fagotto1994b}
---. 1994{\natexlab{b}}, \aaps, 105

\bibitem[{{Falc{\'o}n-Barroso} {et~al.}(2011){Falc{\'o}n-Barroso},
  {S{\'a}nchez-Bl{\'a}zquez}, {Vazdekis}, {Ricciardelli}, {Cardiel}, {Cenarro},
  {Gorgas}, \& {Peletier}}]{FalconBarroso2011}
{Falc{\'o}n-Barroso}, J., {S{\'a}nchez-Bl{\'a}zquez}, P., {Vazdekis}, A.,
  {et~al.} 2011, \aap, 532, A95

\bibitem[{{Fierlinger} {et~al.}(2016){Fierlinger}, {Burkert}, {Ntormousi},
  {Fierlinger}, {Schartmann}, {Ballone}, {Krause}, \& {Diehl}}]{Fierlinger2016}
{Fierlinger}, K.~M., {Burkert}, A., {Ntormousi}, E., {et~al.} 2016, \mnras,
  456, 710

\bibitem[{{Finkelstein} {et~al.}(2012){Finkelstein}, {Papovich}, {Ryan},
  {Pawlik}, {Dickinson}, {Ferguson}, {Finlator}, {Koekemoer}, {Giavalisco},
  {Cooray}, {Dunlop}, {Faber}, {Grogin}, {Kocevski}, \&
  {Newman}}]{Finkelstein2012}
{Finkelstein}, S.~L., {Papovich}, C., {Ryan}, R.~E., {et~al.} 2012, \apj, 758,
  93

\bibitem[{{Freytag} {et~al.}(1996){Freytag}, {Ludwig}, \&
  {Steffen}}]{Freytag1996}
{Freytag}, B., {Ludwig}, H.-G., \& {Steffen}, M. 1996, \aap, 313, 497

\bibitem[{{Fryer}(1999)}]{Fryer1999}
{Fryer}, C.~L. 1999, \apj, 522, 413

\bibitem[{{Fuller} {et~al.}(2015){Fuller}, {Cantiello}, {Lecoanet}, \&
  {Quataert}}]{Fuller2015}
{Fuller}, J., {Cantiello}, M., {Lecoanet}, D., \& {Quataert}, E. 2015, \apj,
  810, 101

\bibitem[{{Gallet} \& {Bouvier}(2013)}]{Gallet2013}
{Gallet}, F., \& {Bouvier}, J. 2013, \aap, 556, A36

\bibitem[{{Georgy} {et~al.}(2013){Georgy}, {Ekstr{\"o}m}, {Eggenberger},
  {Meynet}, {Haemmerl{\'e}}, {Maeder}, {Granada}, {Groh}, {Hirschi}, {Mowlavi},
  {Yusof}, {Charbonnel}, {Decressin}, \& {Barblan}}]{Georgy2013}
{Georgy}, C., {Ekstr{\"o}m}, S., {Eggenberger}, P., {et~al.} 2013, \aap, 558,
  A103

\bibitem[{{Girardi} {et~al.}(2000){Girardi}, {Bressan}, {Bertelli}, \&
  {Chiosi}}]{Girardi2000}
{Girardi}, L., {Bressan}, A., {Bertelli}, G., \& {Chiosi}, C. 2000, \aaps, 141,
  371

\bibitem[{{Gnedin}(2000)}]{Gnedin2000}
{Gnedin}, N.~Y. 2000, \apj, 535, 530

\bibitem[{{Gnedin} \& {Kaurov}(2014)}]{Gnedin2014}
{Gnedin}, N.~Y., \& {Kaurov}, A.~A. 2014, \apj, 793, 30

\bibitem[{{Gnedin} {et~al.}(2008){Gnedin}, {Kravtsov}, \& {Chen}}]{Gnedin2008}
{Gnedin}, N.~Y., {Kravtsov}, A.~V., \& {Chen}, H.-W. 2008, \apj, 672, 765

\bibitem[{{Haardt} \& {Madau}(1996)}]{Haardt1996}
{Haardt}, F., \& {Madau}, P. 1996, \apj, 461, 20

\bibitem[{{Haardt} \& {Madau}(2012)}]{Haardt2012}
---. 2012, \apj, 746, 125

\bibitem[{{Haehnelt}(1995)}]{Haehnelt1995}
{Haehnelt}, M.~G. 1995, \mnras, 273, 249

\bibitem[{{Heger} {et~al.}(2000){Heger}, {Langer}, \& {Woosley}}]{Heger2000}
{Heger}, A., {Langer}, N., \& {Woosley}, S.~E. 2000, \apj, 528, 368

\bibitem[{{Henyey} {et~al.}(1965){Henyey}, {Vardya}, \&
  {Bodenheimer}}]{Henyey1965}
{Henyey}, L., {Vardya}, M.~S., \& {Bodenheimer}, P. 1965, \apj, 142, 841

\bibitem[{{Herwig}(2000)}]{Herwig2000}
{Herwig}, F. 2000, \aap, 360, 952

\bibitem[{{Hopkins}(2013)}]{Hopkins2013}
{Hopkins}, P.~F. 2013, \mnras, 433, 170

\bibitem[{{Hopkins} {et~al.}(2014){Hopkins}, {Kere{\v s}}, {O{\~n}orbe},
  {Faucher-Gigu{\`e}re}, {Quataert}, {Murray}, \& {Bullock}}]{Hopkins2014}
{Hopkins}, P.~F., {Kere{\v s}}, D., {O{\~n}orbe}, J., {et~al.} 2014, \mnras,
  445, 581

\bibitem[{{Hopkins} {et~al.}(2011){Hopkins}, {Quataert}, \&
  {Murray}}]{Hopkins2011}
{Hopkins}, P.~F., {Quataert}, E., \& {Murray}, N. 2011, \mnras, 417, 950

\bibitem[{{Hopkins} {et~al.}(2012){Hopkins}, {Quataert}, \&
  {Murray}}]{Hopkins2012}
---. 2012, \mnras, 421, 3488

\bibitem[{{Huang} {et~al.}(2010){Huang}, {Gies}, \& {McSwain}}]{Huang2010}
{Huang}, W., {Gies}, D.~R., \& {McSwain}, M.~V. 2010, \apj, 722, 605

\bibitem[{{Iwata} {et~al.}(2009){Iwata}, {Inoue}, {Matsuda}, {Furusawa},
  {Hayashino}, {Kousai}, {Akiyama}, {Yamada}, {Burgarella}, \&
  {Deharveng}}]{Iwata2009}
{Iwata}, I., {Inoue}, A.~K., {Matsuda}, Y., {et~al.} 2009, \apj, 692, 1287

\bibitem[{{Joung} \& {Mac Low}(2006)}]{Joung2006}
{Joung}, M.~K.~R., \& {Mac Low}, M.-M. 2006, \apj, 653, 1266

\bibitem[{{Kimm} \& {Cen}(2014)}]{Kimm2014}
{Kimm}, T., \& {Cen}, R. 2014, \apj, 788, 121

\bibitem[{{Kobulnicky} {et~al.}(2014){Kobulnicky}, {Kiminki}, {Lundquist},
  {Burke}, {Chapman}, {Keller}, {Lester}, {Rolen}, {Topel}, {Bhattacharjee},
  {Smullen}, {Vargas {\'A}lvarez}, {Runnoe}, {Dale}, \&
  {Brotherton}}]{Kobulnicky2014}
{Kobulnicky}, H.~A., {Kiminki}, D.~C., {Lundquist}, M.~J., {et~al.} 2014,
  \apjs, 213, 34

\bibitem[{{K{\"o}hler} {et~al.}(2015){K{\"o}hler}, {Langer}, {de Koter}, {de
  Mink}, {Crowther}, {Evans}, {Gr{\"a}fener}, {Sana}, {Sanyal}, {Schneider}, \&
  {Vink}}]{Koehler2015}
{K{\"o}hler}, K., {Langer}, N., {de Koter}, A., {et~al.} 2015, \aap, 573, A71

\bibitem[{{Kroupa}(2001)}]{Kroupa2001}
{Kroupa}, P. 2001, \mnras, 322, 231

\bibitem[{{Kroupa} {et~al.}(2013){Kroupa}, {Weidner}, {Pflamm-Altenburg},
  {Thies}, {Dabringhausen}, {Marks}, \& {Maschberger}}]{Kroupa2013}
{Kroupa}, P., {Weidner}, C., {Pflamm-Altenburg}, J., {et~al.} 2013, {The
  Stellar and Sub-Stellar Initial Mass Function of Simple and Composite
  Populations}, ed. T.~D. {Oswalt} \& G.~{Gilmore}, 115

\bibitem[{{Krumholz} {et~al.}(2012){Krumholz}, {Dekel}, \&
  {McKee}}]{Krumholz2012a}
{Krumholz}, M.~R., {Dekel}, A., \& {McKee}, C.~F. 2012, \apj, 745, 69

\bibitem[{{Krumholz} {et~al.}(2011){Krumholz}, {Klein}, \&
  {McKee}}]{Krumholz2011}
{Krumholz}, M.~R., {Klein}, R.~I., \& {McKee}, C.~F. 2011, \apj, 740, 74

\bibitem[{{Krumholz} \& {Matzner}(2009)}]{Krumholz2009}
{Krumholz}, M.~R., \& {Matzner}, C.~D. 2009, \apj, 703, 1352

\bibitem[{{Krumholz} \& {Thompson}(2012)}]{Krumholz2012}
{Krumholz}, M.~R., \& {Thompson}, T.~A. 2012, \apj, 760, 155

\bibitem[{{Langer}(1992)}]{Langer1992}
{Langer}, N. 1992, \aap, 265, L17

\bibitem[{{Langer}(1998)}]{Langer1998}
---. 1998, \aap, 329, 551

\bibitem[{{Leitet} {et~al.}(2013){Leitet}, {Bergvall}, {Hayes}, {Linn{\'e}}, \&
  {Zackrisson}}]{Leitet2013}
{Leitet}, E., {Bergvall}, N., {Hayes}, M., {Linn{\'e}}, S., \& {Zackrisson}, E.
  2013, \aap, 553, A106

\bibitem[{{Leitherer} {et~al.}(2014){Leitherer}, {Ekstr{\"o}m}, {Meynet},
  {Schaerer}, {Agienko}, \& {Levesque}}]{Leitherer2014}
{Leitherer}, C., {Ekstr{\"o}m}, S., {Meynet}, G., {et~al.} 2014, \apjs, 212, 14

\bibitem[{{Leitherer} {et~al.}(2010){Leitherer}, {Ortiz Ot{\'a}lvaro},
  {Bresolin}, {Kudritzki}, {Lo Faro}, {Pauldrach}, {Pettini}, \&
  {Rix}}]{Leitherer2010}
{Leitherer}, C., {Ortiz Ot{\'a}lvaro}, P.~A., {Bresolin}, F., {et~al.} 2010,
  \apjs, 189, 309

\bibitem[{{Leitherer} {et~al.}(1992){Leitherer}, {Robert}, \&
  {Drissen}}]{Leitherer1992}
{Leitherer}, C., {Robert}, C., \& {Drissen}, L. 1992, \apj, 401, 596

\bibitem[{{Leitherer} {et~al.}(1999){Leitherer}, {Schaerer}, {Goldader},
  {Delgado}, {Robert}, {Kune}, {de Mello}, {Devost}, \&
  {Heckman}}]{Leitherer1999}
{Leitherer}, C., {Schaerer}, D., {Goldader}, J.~D., {et~al.} 1999, \apjs, 123,
  3

\bibitem[{{Levesque} {et~al.}(2012){Levesque}, {Leitherer}, {Ekstrom},
  {Meynet}, \& {Schaerer}}]{Levesque2012}
{Levesque}, E.~M., {Leitherer}, C., {Ekstrom}, S., {Meynet}, G., \& {Schaerer},
  D. 2012, \apj, 751, 67

\bibitem[{{Lopez} {et~al.}(2014){Lopez}, {Krumholz}, {Bolatto}, {Prochaska},
  {Ramirez-Ruiz}, \& {Castro}}]{Lopez2014}
{Lopez}, L.~A., {Krumholz}, M.~R., {Bolatto}, A.~D., {et~al.} 2014, \apj, 795,
  121

\bibitem[{{Lucy} \& {Solomon}(1970)}]{Lucy1970}
{Lucy}, L.~B., \& {Solomon}, P.~M. 1970, \apj, 159, 879

\bibitem[{{Ma} {et~al.}(2016{\natexlab{a}}){Ma}, {Hopkins},
  {Faucher-Gigu{\`e}re}, {Zolman}, {Muratov}, {Kere{\v s}}, \&
  {Quataert}}]{Ma2016a}
{Ma}, X., {Hopkins}, P.~F., {Faucher-Gigu{\`e}re}, C.-A., {et~al.}
  2016{\natexlab{a}}, \mnras, 456, 2140

\bibitem[{{Ma} {et~al.}(2016{\natexlab{b}}){Ma}, {Hopkins}, {Kasen},
  {Quataert}, {Faucher-Gigu{\`e}re}, {Kere{\v s}}, {Murray}, \&
  {Strom}}]{Ma2016b}
{Ma}, X., {Hopkins}, P.~F., {Kasen}, D., {et~al.} 2016{\natexlab{b}}, \mnras,
  459, 3614

\bibitem[{{Ma} {et~al.}(2015){Ma}, {Kasen}, {Hopkins}, {Faucher-Gigu{\`e}re},
  {Quataert}, {Kere{\v s}}, \& {Murray}}]{Ma2015}
{Ma}, X., {Kasen}, D., {Hopkins}, P.~F., {et~al.} 2015, \mnras, 453, 960

\bibitem[{{Mac Low} \& {Klessen}(2004)}]{MacLow2004}
{Mac Low}, M.-M., \& {Klessen}, R.~S. 2004, Reviews of Modern Physics, 76, 125

\bibitem[{{Madau} {et~al.}(1999){Madau}, {Haardt}, \& {Rees}}]{Madau1999}
{Madau}, P., {Haardt}, F., \& {Rees}, M.~J. 1999, \apj, 514, 648

\bibitem[{{Maeder}(1987)}]{Maeder1987}
{Maeder}, A. 1987, \aap, 178, 159

\bibitem[{{Maeder}(1990)}]{Maeder1990}
---. 1990, \aaps, 84, 139

\bibitem[{{Maeder} \& {Meynet}(2000)}]{Maeder2000}
{Maeder}, A., \& {Meynet}, G. 2000, \araa, 38, 143

\bibitem[{{Magic} {et~al.}(2010){Magic}, {Serenelli}, {Weiss}, \&
  {Chaboyer}}]{Magic2010}
{Magic}, Z., {Serenelli}, A., {Weiss}, A., \& {Chaboyer}, B. 2010, \apj, 718,
  1378

\bibitem[{{Martin}(1999)}]{Martin1999}
{Martin}, C.~L. 1999, \apj, 513, 156

\bibitem[{{Martins} {et~al.}(2013){Martins}, {Depagne}, {Russeil}, \&
  {Mahy}}]{Martins2013}
{Martins}, F., {Depagne}, E., {Russeil}, D., \& {Mahy}, L. 2013, \aap, 554, A23

\bibitem[{{Matzner}(2002)}]{Matzner2002}
{Matzner}, C.~D. 2002, \apj, 566, 302

\bibitem[{{McKee} \& {Ostriker}(2007)}]{McKee2007}
{McKee}, C.~F., \& {Ostriker}, E.~C. 2007, \araa, 45, 565

\bibitem[{{McKee} \& {Ostriker}(1977)}]{McKee1977}
{McKee}, C.~F., \& {Ostriker}, J.~P. 1977, \apj, 218, 148

\bibitem[{{Meynet} {et~al.}(1994){Meynet}, {Maeder}, {Schaller}, {Schaerer}, \&
  {Charbonnel}}]{Meynet1994}
{Meynet}, G., {Maeder}, A., {Schaller}, G., {Schaerer}, D., \& {Charbonnel}, C.
  1994, \aaps, 103

\bibitem[{{Monreal-Ibero} {et~al.}(2012){Monreal-Ibero}, {Walsh}, \&
  {V{\'{\i}}lchez}}]{MonrealIbero2012}
{Monreal-Ibero}, A., {Walsh}, J.~R., \& {V{\'{\i}}lchez}, J.~M. 2012, \aap,
  544, A60

\bibitem[{{Muratov} {et~al.}(2015){Muratov}, {Kere{\v s}},
  {Faucher-Gigu{\`e}re}, {Hopkins}, {Quataert}, \& {Murray}}]{Muratov2015}
{Muratov}, A.~L., {Kere{\v s}}, D., {Faucher-Gigu{\`e}re}, C.-A., {et~al.}
  2015, \mnras, 454, 2691

\bibitem[{{Murray} {et~al.}(2011){Murray}, {M{\'e}nard}, \&
  {Thompson}}]{Murray2011}
{Murray}, N., {M{\'e}nard}, B., \& {Thompson}, T.~A. 2011, \apj, 735, 66

\bibitem[{{Murray} {et~al.}(2005){Murray}, {Quataert}, \&
  {Thompson}}]{Murray2005}
{Murray}, N., {Quataert}, E., \& {Thompson}, T.~A. 2005, \apj, 618, 569

\bibitem[{{Murray} {et~al.}(2010){Murray}, {Quataert}, \&
  {Thompson}}]{Murray2010}
---. 2010, \apj, 709, 191

\bibitem[{{Narayanan} \& {Dav{\'e}}(2013)}]{Narayanan2013}
{Narayanan}, D., \& {Dav{\'e}}, R. 2013, \mnras, 436, 2892

\bibitem[{{Nath} \& {Silk}(2009)}]{Nath2009}
{Nath}, B.~B., \& {Silk}, J. 2009, \mnras, 396, L90

\bibitem[{{Nugis} \& {Lamers}(2000)}]{Nugis2000}
{Nugis}, T., \& {Lamers}, H.~J.~G.~L.~M. 2000, \aap, 360, 227

\bibitem[{{Oppenheimer} \& {Dav{\'e}}(2006)}]{Oppenheimer2006}
{Oppenheimer}, B.~D., \& {Dav{\'e}}, R. 2006, \mnras, 373, 1265

\bibitem[{{Pauldrach}(2012)}]{Pauldrach2012}
{Pauldrach}, A.~W.~A. 2012, {WM-basic: Modeling atmospheres of hot stars},
  Astrophysics Source Code Library, ascl:1204.001

\bibitem[{{Paxton} {et~al.}(2011){Paxton}, {Bildsten}, {Dotter}, {Herwig},
  {Lesaffre}, \& {Timmes}}]{Paxton2011}
{Paxton}, B., {Bildsten}, L., {Dotter}, A., {et~al.} 2011, \apjs, 192, 3

\bibitem[{{Paxton} {et~al.}(2013){Paxton}, {Cantiello}, {Arras}, {Bildsten},
  {Brown}, {Dotter}, {Mankovich}, {Montgomery}, {Stello}, {Timmes}, \&
  {Townsend}}]{Paxton2013}
{Paxton}, B., {Cantiello}, M., {Arras}, P., {et~al.} 2013, \apjs, 208, 4

\bibitem[{{Paxton} {et~al.}(2015){Paxton}, {Marchant}, {Schwab}, {Bauer},
  {Bildsten}, {Cantiello}, {Dessart}, {Farmer}, {Hu}, {Langer}, {Townsend},
  {Townsley}, \& {Timmes}}]{Paxton2015}
{Paxton}, B., {Marchant}, P., {Schwab}, J., {et~al.} 2015, \apjs, 220, 15

\bibitem[{{Penny}(1996)}]{Penny1996}
{Penny}, L.~R. 1996, \apj, 463, 737

\bibitem[{{Penny} \& {Gies}(2009)}]{Penny2009}
{Penny}, L.~R., \& {Gies}, D.~R. 2009, \apj, 700, 844

\bibitem[{{Pinsonneault} {et~al.}(1989){Pinsonneault}, {Kawaler}, {Sofia}, \&
  {Demarque}}]{Pinsonneault1989}
{Pinsonneault}, M.~H., {Kawaler}, S.~D., {Sofia}, S., \& {Demarque}, P. 1989,
  \apj, 338, 424

\bibitem[{{Planck Collaboration} {et~al.}(2015){Planck Collaboration}, {Ade},
  {Aghanim}, {Arnaud}, {Ashdown}, {Aumont}, {Baccigalupi}, {Banday},
  {Barreiro}, {Bartlett}, \& et~al.}]{Planck2015}
{Planck Collaboration}, {Ade}, P.~A.~R., {Aghanim}, N., {et~al.} 2015, ArXiv
  e-prints, arXiv:1502.01589

\bibitem[{{Potter} {et~al.}(2012){Potter}, {Tout}, \& {Eldridge}}]{Potter2012a}
{Potter}, A.~T., {Tout}, C.~A., \& {Eldridge}, J.~J. 2012, \mnras, 419, 748

\bibitem[{{Ram{\'{\i}}rez-Agudelo} {et~al.}(2013){Ram{\'{\i}}rez-Agudelo},
  {Sim{\'o}n-D{\'{\i}}az}, {Sana}, {de Koter}, {Sab{\'{\i}}n-Sanjul{\'{\i}}an},
  {de Mink}, {Dufton}, {Gr{\"a}fener}, {Evans}, {Herrero}, {Langer}, {Lennon},
  {Ma{\'{\i}}z Apell{\'a}niz}, {Markova}, {Najarro}, {Puls}, {Taylor}, \&
  {Vink}}]{RamirezAgudelo2013}
{Ram{\'{\i}}rez-Agudelo}, O.~H., {Sim{\'o}n-D{\'{\i}}az}, S., {Sana}, H.,
  {et~al.} 2013, \aap, 560, A29

\bibitem[{{Razoumov} \& {Sommer-Larsen}(2010)}]{Razoumov2010}
{Razoumov}, A.~O., \& {Sommer-Larsen}, J. 2010, \apj, 710, 1239

\bibitem[{{Robertson} {et~al.}(2013){Robertson}, {Furlanetto}, {Schneider},
  {Charlot}, {Ellis}, {Stark}, {McLure}, {Dunlop}, {Koekemoer}, {Schenker},
  {Ouchi}, {Ono}, {Curtis-Lake}, {Rogers}, {Bowler}, \&
  {Cirasuolo}}]{Robertson2013}
{Robertson}, B.~E., {Furlanetto}, S.~R., {Schneider}, E., {et~al.} 2013, \apj,
  768, 71

\bibitem[{{Salpeter}(1955)}]{Salpeter1955}
{Salpeter}, E.~E. 1955, \apj, 121, 161

\bibitem[{{Sana} {et~al.}(2012){Sana}, {de Mink}, {de Koter}, {Langer},
  {Evans}, {Gieles}, {Gosset}, {Izzard}, {Le Bouquin}, \&
  {Schneider}}]{Sana2012}
{Sana}, H., {de Mink}, S.~E., {de Koter}, A., {et~al.} 2012, Science, 337, 444

\bibitem[{{Sana} {et~al.}(2013){Sana}, {de Koter}, {de Mink}, {Dunstall},
  {Evans}, {H{\'e}nault-Brunet}, {Ma{\'{\i}}z Apell{\'a}niz},
  {Ram{\'{\i}}rez-Agudelo}, {Taylor}, {Walborn}, {Clark}, {Crowther},
  {Herrero}, {Gieles}, {Langer}, {Lennon}, \& {Vink}}]{Sana2013}
{Sana}, H., {de Koter}, A., {de Mink}, S.~E., {et~al.} 2013, \aap, 550, A107

\bibitem[{{S{\'a}nchez-Bl{\'a}zquez} {et~al.}(2006){S{\'a}nchez-Bl{\'a}zquez},
  {Peletier}, {Jim{\'e}nez-Vicente}, {Cardiel}, {Cenarro},
  {Falc{\'o}n-Barroso}, {Gorgas}, {Selam}, \& {Vazdekis}}]{SanchezBlazquez2006}
{S{\'a}nchez-Bl{\'a}zquez}, P., {Peletier}, R.~F., {Jim{\'e}nez-Vicente}, J.,
  {et~al.} 2006, \mnras, 371, 703

\bibitem[{{Siana} {et~al.}(2010){Siana}, {Teplitz}, {Ferguson}, {Brown},
  {Giavalisco}, {Dickinson}, {Chary}, {de Mello}, {Conselice}, {Bridge},
  {Gardner}, {Colbert}, \& {Scarlata}}]{Siana2010}
{Siana}, B., {Teplitz}, H.~I., {Ferguson}, H.~C., {et~al.} 2010, \apj, 723, 241

\bibitem[{{Smith} {et~al.}(2016){Smith}, {Crowther}, {Calzetti}, \&
  {Sidoli}}]{Smith2016}
{Smith}, L.~J., {Crowther}, P.~A., {Calzetti}, D., \& {Sidoli}, F. 2016, \apj,
  823, 38

\bibitem[{{Smith} {et~al.}(2002){Smith}, {Norris}, \& {Crowther}}]{Smith2002}
{Smith}, L.~J., {Norris}, R.~P.~F., \& {Crowther}, P.~A. 2002, \mnras, 337,
  1309

\bibitem[{{Song} {et~al.}(2016){Song}, {Meynet}, {Maeder}, {Ekstr{\"o}m}, \&
  {Eggenberger}}]{Song2016}
{Song}, H.~F., {Meynet}, G., {Maeder}, A., {Ekstr{\"o}m}, S., \& {Eggenberger},
  P. 2016, \aap, 585, A120

\bibitem[{{Spruit}(2002)}]{Spruit2002}
{Spruit}, H.~C. 2002, \aap, 381, 923

\bibitem[{{Stanway} {et~al.}(2016){Stanway}, {Eldridge}, \&
  {Becker}}]{Stanway2016}
{Stanway}, E.~R., {Eldridge}, J.~J., \& {Becker}, G.~D. 2016, \mnras, 456, 485

\bibitem[{{Steidel} {et~al.}(2016){Steidel}, {Strom}, {Pettini}, {Rudie},
  {Reddy}, \& {Trainor}}]{Steidel2016}
{Steidel}, C.~C., {Strom}, A.~L., {Pettini}, M., {et~al.} 2016, \apj, 826, 159

\bibitem[{{Steidel} {et~al.}(2014){Steidel}, {Rudie}, {Strom}, {Pettini},
  {Reddy}, {Shapley}, {Trainor}, {Erb}, {Turner}, {Konidaris}, {Kulas}, {Mace},
  {Matthews}, \& {McLean}}]{Steidel2014}
{Steidel}, C.~C., {Rudie}, G.~C., {Strom}, A.~L., {et~al.} 2014, \apj, 795, 165

\bibitem[{{Sukhbold} \& {Woosley}(2014)}]{Sukhbold2014}
{Sukhbold}, T., \& {Woosley}, S.~E. 2014, \apj, 783, 10

\bibitem[{{Tamburro} {et~al.}(2009){Tamburro}, {Rix}, {Leroy}, {Mac Low},
  {Walter}, {Kennicutt}, {Brinks}, \& {de Blok}}]{Tamburro2009}
{Tamburro}, D., {Rix}, H.-W., {Leroy}, A.~K., {et~al.} 2009, \aj, 137, 4424

\bibitem[{{Turner} {et~al.}(2015){Turner}, {Beck}, {Benford}, {Consiglio},
  {Ho}, {Kov{\'a}cs}, {Meier}, \& {Zhao}}]{Turner2015}
{Turner}, J.~L., {Beck}, S.~C., {Benford}, D.~J., {et~al.} 2015, \nat, 519, 331

\bibitem[{{V{\'a}zquez} \& {Leitherer}(2005)}]{Vazquez2005}
{V{\'a}zquez}, G.~A., \& {Leitherer}, C. 2005, \apj, 621, 695

\bibitem[{{Villaverde} {et~al.}(2010){Villaverde}, {Cervi{\~n}o}, \&
  {Luridiana}}]{Villaverde2010}
{Villaverde}, M., {Cervi{\~n}o}, M., \& {Luridiana}, V. 2010, \aap, 522, A49

\bibitem[{{Vink} {et~al.}(2000){Vink}, {de Koter}, \& {Lamers}}]{Vink2000}
{Vink}, J.~S., {de Koter}, A., \& {Lamers}, H.~J.~G.~L.~M. 2000, \aap, 362, 295

\bibitem[{{Vink} {et~al.}(2001){Vink}, {de Koter}, \& {Lamers}}]{Vink2001}
---. 2001, \aap, 369, 574

\bibitem[{{Walch} {et~al.}(2012){Walch}, {Whitworth}, {Bisbas}, {W{\"u}nsch},
  \& {Hubber}}]{Walch2012}
{Walch}, S.~K., {Whitworth}, A.~P., {Bisbas}, T., {W{\"u}nsch}, R., \&
  {Hubber}, D. 2012, \mnras, 427, 625

\bibitem[{{Weisz} {et~al.}(2015){Weisz}, {Johnson}, {Foreman-Mackey},
  {Dolphin}, {Beerman}, {Williams}, {Dalcanton}, {Rix}, {Hogg}, {Fouesneau},
  {Johnson}, {Bell}, {Boyer}, {Gouliermis}, {Guhathakurta}, {Kalirai}, {Lewis},
  {Seth}, \& {Skillman}}]{Weisz2015}
{Weisz}, D.~R., {Johnson}, L.~C., {Foreman-Mackey}, D., {et~al.} 2015, \apj,
  806, 198

\bibitem[{{Whitworth}(1979)}]{Whitworth1979}
{Whitworth}, A. 1979, \mnras, 186, 59

\bibitem[{{Williams} \& {McKee}(1997)}]{Williams1997}
{Williams}, J.~P., \& {McKee}, C.~F. 1997, \apj, 476, 166

\bibitem[{{Wise} {et~al.}(2014){Wise}, {Demchenko}, {Halicek}, {Norman},
  {Turk}, {Abel}, \& {Smith}}]{Wise2014}
{Wise}, J.~H., {Demchenko}, V.~G., {Halicek}, M.~T., {et~al.} 2014, \mnras,
  442, 2560

\bibitem[{{Wolff} {et~al.}(2004){Wolff}, {Strom}, \& {Hillenbrand}}]{Wolff2004}
{Wolff}, S.~C., {Strom}, S.~E., \& {Hillenbrand}, L.~A. 2004, \apj, 601, 979

\bibitem[{{Woosley} \& {Heger}(2006)}]{Woosley2006}
{Woosley}, S.~E., \& {Heger}, A. 2006, \apj, 637, 914

\bibitem[{{Yoon} \& {Langer}(2005)}]{Yoon2005}
{Yoon}, S.-C., \& {Langer}, N. 2005, \aap, 443, 643

\bibitem[{{Yusof} {et~al.}(2013){Yusof}, {Hirschi}, {Meynet}, {Crowther},
  {Ekstr{\"o}m}, {Frischknecht}, {Georgy}, {Abu Kassim}, \&
  {Schnurr}}]{Yusof2013}
{Yusof}, N., {Hirschi}, R., {Meynet}, G., {et~al.} 2013, \mnras, 433, 1114

\bibitem[{{Zahn}(1983)}]{Zahn1983}
{Zahn}, J.-P. 1983, in Saas-Fee Advanced Course 13: Astrophysical Processes in
  Upper Main Sequence Stars, ed. A.~N. {Cox}, S.~{Vauclair}, \& J.~P. {Zahn},
  253

\bibitem[{{Zinnecker} \& {Yorke}(2007)}]{Zinnecker2007}
{Zinnecker}, H., \& {Yorke}, H.~W. 2007, \araa, 45, 481

\end{thebibliography}

\end{document}